\documentclass[lettersize,journal]{IEEEtran}
\usepackage{amsmath,amsfonts}
\usepackage{algorithmic}
\usepackage{algorithm}
\usepackage{array}
\usepackage{textcomp}
\usepackage{stfloats}
\usepackage{url}
\usepackage{verbatim}
\usepackage{graphicx}
\usepackage{cite}
\usepackage{bm}
\usepackage{subfigure}

\begin{document}

\title{A Blockchain-empowered Multi-Aggregator Federated Learning Architecture in Edge Computing with Deep Reinforcement Learning Optimization}

\author{Xiao Li,~\IEEEmembership{Student Member,~IEEE,} and Weili Wu,~\IEEEmembership{Senior Member,~IEEE,}
\thanks{X. Li and W. Wu are with Department of Computer Science at The University of Texas at Dallas.}
\thanks{Coresponding Author: Xiao Li (Xiao.Li@utdallas.edu)}
\thanks{This work is partially supported by NSF Grant No. 1822985 and No. 1907472}}



\maketitle

\begin{abstract}
Federated learning (FL) is emerging as a sought-after distributed machine learning architecture, offering the advantage of model training without direct exposure of raw data. With advancements in network infrastructure, FL has been seamlessly integrated into edge computing. However, the limited resources on edge devices introduce security vulnerabilities to FL in the context. While blockchain technology promises to bolster security, practical deployment on resource-constrained edge devices remains a challenge. Moreover, the exploration of FL with multiple aggregators in edge computing is still new in the literature. Addressing these gaps, we introduce the Blockchain-empowered Heterogeneous Multi-Aggregator Federated Learning Architecture (BMA-FL). We design a novel light-weight Byzantine consensus mechanism, namely PBCM, to enable secure and fast model aggregation and synchronization in BMA-FL. We also dive into the heterogeneity problem in BMA-FL that the aggregators are associated with varied number of connected trainers with Non-IID data distributions and diverse training speed. We proposed a multi-agent deep reinforcement learning algorithm to help aggregators decide the best training strategies. The experiments on real-word datasets demonstrate the efficiency of BMA-FL to achieve better models faster than baselines, showing the efficacy of PBCM and proposed deep reinforcement learning algorithm. 

\end{abstract}


\section{Introduction}

\IEEEPARstart{W}{ith} the rapid development of internet of things (IoT), edge devices such as mobile devices, smart sensor and smart meters are enhancing the productivity and efficiency of data collecting~\cite{DBLP:journals/iotj/SanghamiLH23,DBLP:journals/iotj/NguyenDPPLSLNP21}, which facilitates the development of intelligent future. Machine learning models are widely used in intelligent applications to obtaining insights from enormous data on edge devices~\cite{DBLP:journals/comsur/NguyenCDLPLSLP21}. 

However, the abuse of intelligent applications brings recent privacy concerns from public about the unrestrained data collection. 
European Union has announced multiple regulations to restrict the private data exposure and acquisition~\cite{DBLP:journals/iotj/LoLLWXPZ23}. 
Private data should not be shared with third parties, and transmitting sensitive information over the internet could expose it to cybersecurity threats.

Federated Learning (FL)~\cite{DBLP:conf/aistats/McMahanMRHA17} is investigated and becoming a popular solution that allows machine learning model requesters/owners get model trained without having to collecting data from edge devices at unintended cost and risks. In a typical federated system, a model is trained across multiple trainers who have the training data. An aggregator will collect and aggregate the trained models. Then the aggregated model is sent back to trainers for another round of training. As the the data on each trainer are never shared, the privacy can be preserved. 

The distributed architecture of FL brings trust issues among the model requester and trainers. Malicious trainers may upload random models to spoil the global model, and lead to failure of training~\cite{DBLP:journals/kbs/ZhaoZC23,DBLP:conf/aaai/YanWYL23}. Specially, in edge computing context, edge devices introduce security vulnerabilities to the model training because of limited supervision and low security robustness~\cite{DBLP:journals/iotj/ZhangGQXQW23}. 
Incorporating blockchain modules in to FL phases can be helpful to validate the models without having to build a trustable third-party to do authentication. The smart contract deployed on the blockchain system can ensure the quality of model updates, avoiding malicious trainers that attempts to damage the global model. The tractability of blockchain storage also provides transparent model update and training log to all participants~\cite{DBLP:journals/tpds/ShayanFYB21, DBLP:journals/tii/LuHDMZ20a, DBLP:journals/tpds/LiSWDMSHP22, DBLP:journals/tii/KalapaakingKRAY23, DBLP:journals/tpds/WangHLXX23}. 

In most Blockchain-empowered FL (BFL) work in edge computing, edge devices are often set as the peers in blockchain modules based on intuition that edge devices are trainers and thus have to share parameters via blockchain. However, the blockchain module imposes extra computation and communication burden to edge devices while edge devices, e.g. micro computers or sensors have limited computation power. Therefore it is not practical to involve those edge devices into blockchain modules such as~\cite{DBLP:journals/tpds/WangHLXX23, DBLP:journals/ieeenl/HieuTNNKE22, DBLP:journals/tits/AloqailyRG22} did in their work. Especially when Proof of Work (PoW) consensus mechanism is used as the consensus mechanism in the blockchain module~\cite{DBLP:journals/tpds/LiSWDMSHP22, DBLP:conf/bigdataconf/ChenJLLL18, DBLP:journals/iotj/QuGLXYLZ20}, tremendous computation costs are applied to edge devices, who are supposed to focus resources on model training instead of solving meaningless hash puzzles in PoW.

Moreover, most BFL framework are proposed under the context where only one aggregator exits which can not reflect the common real-word use cases~\cite{DBLP:journals/iotj/LoLLWXPZ23, DBLP:journals/icl/KimPBK20}. In large-scale edge computing systems, where edge devices are broadly distributed in wide range of spaces, edge devices are usually connected through multiple edge servers, such as network access points or base stations. 

In this paper, we address these under-explored problems by proposing a novel architecture of blockchain-FL framework in edge computing context where multiple edge servers are serving as multiple aggregators. 

\begin{figure}[thb]
    \centering
    \includegraphics[width = 0.8\linewidth]{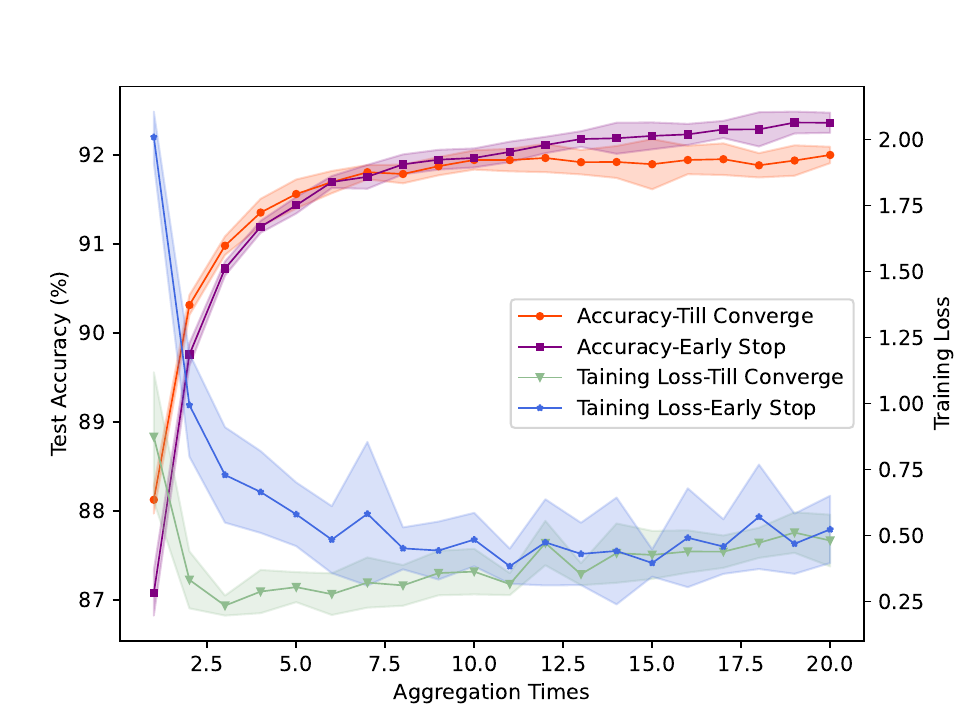}
    \caption{The Training Process Comparison between ``Till Converge" and ``Early Stop" }
    \label{fig:SimpComp}
\end{figure}

In addition, due to the heterogeneity of data holding by edge devices, the models converged on local data will lost generalizability which will ruin the performance of aggregated models on unseen data. We design a simple demonstrative experiment to illustrate the problem. We create an aggregator with 30 trainers holding FashionMnist dataset in heterogeneous distribution. The conventional way for FL training is that each trainer trains the model on their local data till convergence. We compare this strategy with an early stopping strategy that all trainers will do at most 10 epochs of training. As shown in Table~\ref{tab:comp_early_stop}, ``early stop" strategy finished 20 rounds aggregation with 36\% less time, but achieved better final performance. By stop the training after 10 epochs on each trainer, ``early stop" went through 62\% less data samples which saves significant computation resources. Figure~\ref{fig:SimpComp} shows the performance of two strategies after each aggregations. It is crucial for trainers and aggregators try to obtain the best model as usually the reward is associated with the model performance. In this paper, we solve this problem in a more challenging context that multiple aggregators with varied number of trainers exist in BFL.


\begin{table}[htb]
\centering
\caption{An Example of Trainers Behaviour and Corresponding Results}
\label{tab:comp_early_stop}
\begin{tabular}{c|c|c}
\hline
Stats                           & Till Converge & Early Stop \\ \hline
Time Spent                      &  78.67min                 &     49.78min                     \\ \hline
Final Accuracy                  & 91.78\%                &    92.36\%                       \\ \hline
Total \# Data Samples &       29,638,581.11         &         11,018,445.44                \\ \hline
Total \# Aggregation     & 20                     &         20                  \\ \hline
\end{tabular}
\end{table}

The contribution of this paper can be summarized as: 
\begin{enumerate}
    \item We introduce the Blockchain-empowered Heterogeneous Multi-Aggregator Federated Learning architecture (BMA-FL) into edge computing context, where each aggregator is connected with varied number of trainers with Non-IID data and CPU speed distribution. 
    \item We design a blockchain module to enable fast, secure and transparent global model consensus. Specifically, a Performance-based Byzantine Consensus Mechanism is proposed to select a miner for for aggregating models from all aggregators and executing global model updates.
    \item We propose a Multi-Agent with Shared Buffer Deep Reinforcement Learning algorithm (MASB-DRL). This approach empowers aggregators to refine their training strategies, enhancing the speed and performance of the BMA-FL. 
    \item We conduct experiment on common real-world datasets to demonstrate the efficacy and efficiency of proposed BMA-FL architecture and MASB-DRL algorithm.
\end{enumerate}

The remainder of this paper is organized as: In Section~\ref{sec:rel_wor} we review recent related work in literature. In Section~\ref{sec:framework} we describe the proposed multi-aggregator federated learning framework in edge computing. After that the Multi-Agent with Shared Buffer Deep Reinforcement Learning algorithm is presented in Section~\ref{sec:drl}. Then the evaluation of proposed architecture is presented in Section~\ref{sec:eval}. Finally we conclude this paper in Section~\ref{sec:conc}.

\section{Related Work}
\label{sec:rel_wor}
Blockchain system can elevate the security level of data transmission among edge devices due to the vulnerabilities to security attack, also introduce fairness and decentralization to federated learning~\cite{DBLP:journals/csur/IssaMTST23}.
Blockchain recently has been widely introduced into federated learning in edge computing context~\cite{DBLP:journals/cn/WanQGX22, DBLP:journals/iotj/NguyenDPPLSLNP21} in various use cases. Hu et al.~\cite{DBLP:journals/iotj/HuWXC23} and Wang et al.~\cite{DBLP:journals/cn/WangWHMSTC22} design blockchain-based FL frameworks for mobile crowd-sourcing. Otoum et al.~\cite{DBLP:journals/iotj/OtoumRM23} study decentralized sustainable energy trade with security and robustness ensured by blockchain and FL. Other topics such as digital twin network~\cite{DBLP:journals/iotj/JiangZTXZ22,DBLP:journals/iotj/LuHZMZ21}, content cache~\cite{DBLP:journals/iotj/CuiSMCYZX22} and energy storage~\cite{DBLP:journals/peerj-cs/MengL22} are also studied with incorporation of BFL.

\begin{figure}[thb]
    \centering
    \includegraphics[width = 1.0\linewidth]{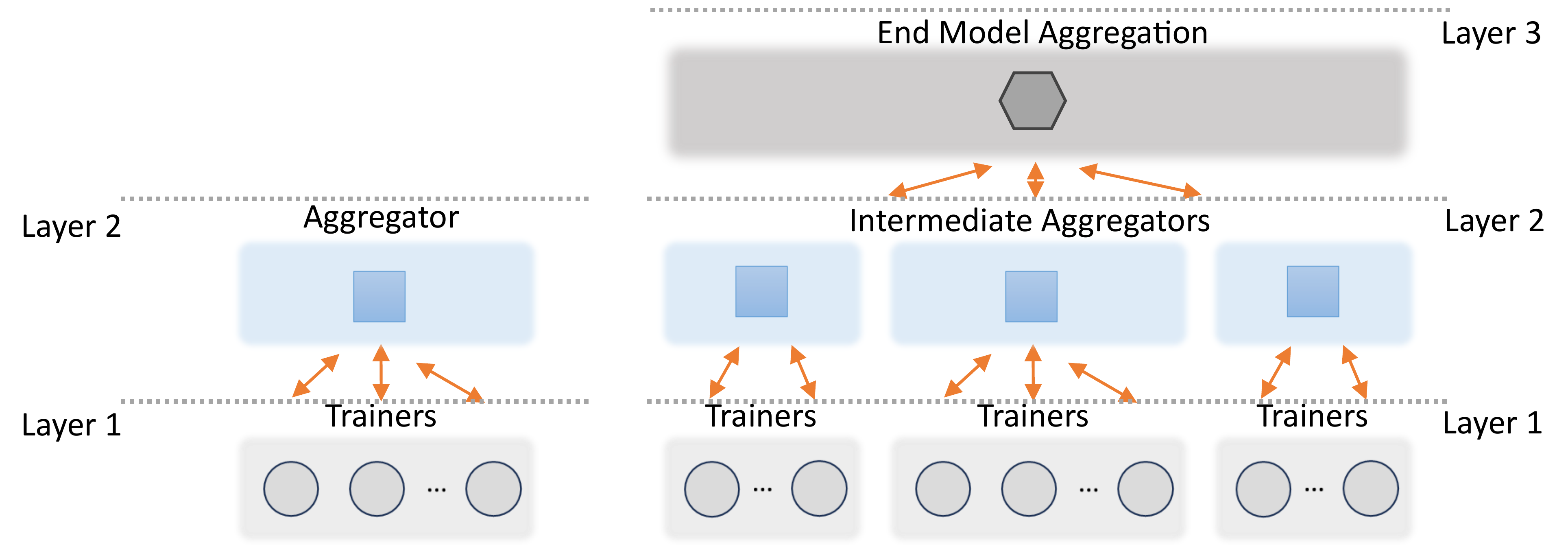}
    \caption{Classic 2-layer FL architecture vs 3-layer FL architecture studied in this paper}
    \label{fig:2vs3}
\end{figure}

Most of existing BFL frameworks in edge computing are composed in two layers i.e. one model owner/aggregator and multiple model trainers as shown in Figure~\ref{fig:2vs3}. Fan et al.~\cite{DBLP:journals/iotj/FanZZC21} considers three layer but the the IoT nodes in lowest layer are only for providing data to edge nodes and does not involve any function in FL training. Cui et al.~\cite{DBLP:journals/iotj/CuiSMCYZX22} proposed similar three layer architecture as in this paper where there are also multiple edge nodes. But these edge nodes are working as trainers and the models are still aggregate by one central server, which is still not a multi-aggregator architecture.
Liu et al.~\cite{DBLP:journals/tvt/LiuZZZSPZ21} utilizes distributed aggregation in vehicular network which allows multiple aggregators in the vehicular network sharing models trained by connected vehicles. But the author did not study the heterogeneous problem among the aggregators and Proof of Accuracy (PoA) will make the blockchain module become centralized very fast. Nguyen et al.~\cite{DBLP:journals/jsac/NguyenHLPB22} also considers multiple edge servers as aggregators similar to this paper, but we do not include edge devices into the blockchain mining to save resources on edge devices. Lee and Kim~\cite{DBLP:journals/sensors/LeeK22b} also study multi-aggregator with heterogeneous data distribution and design  local aggregation and global aggregation phases in their proposed architecture. Though the terms are similar in this paper, FL is not conducted within edge computing scope and we have totally different blockchain design and aggregation formulation. 

Because maintaining blockchain impose extra computation and storage cost to edge devices, researchers have been trying to propose novel consensus mechanisms in edge computing context as substitutes of Proof of Work (PoW) which is current the most common consensus mechanism.  Qu et al.~\cite{DBLP:journals/tpds/QuWHC21} proposing Proof of Federated Learning (PoFL) and Lu et al.~\cite{DBLP:journals/tii/LuHDMZ20a} proposing Proof of Quality (PoQ) are using the training performance as the criteria to select miner. Wang et al.~\cite{DBLP:journals/jsac/WangPSLBW22} propose PF-PoFL as an extension to PoFL, which removes the central platform and forming a dynamically optimized pooled structure for AI model training. 
Li et al.~\cite{DBLP:journals/network/LiCLHZY21} propose Proof of Committee, that a selected committee is responsible for model validation and blockchain generation based on evaluating collected model's performance on the data held by the committee. Chen et al.~\cite{DBLP:journals/iotj/ChenLWYLXZC23} propose delegated Byzantine fault-tolerant consensus that divides all nodes into accounting nodes who will be block minders and ordinary nodes who will vote accounting nodes. The consensus is only reached among the accounting nodes, instead of sending consensus messages to all system nodes which reduces the consensus delay.

Another direction to reduce the blockchain consensus cost is to find the best resource allocation on edge devices so that the blockchain will not consume too much resources cause insufficiency of model training. Reinforcement learning-based algorithm are often taken into account to tackle this issue. Nguyen et al.~\cite{DBLP:journals/ieeenl/HieuTNNKE22} use Deep Reinforcement Learning (DRL) to decide the amount of data and energy used on edge device as well as the blockchain generation speed to reduce the model training latency. Lu et al.~\cite{DBLP:journals/iotj/LuHZMZ21} focus on the communication cost of blockchain system, they designed a DRL algorithm to optimize the communication bandwidth allocation. Lin et al.~\cite{10159403} propose a streamline-based shard transmission mechanism for BFL in transportation system, where DRL is adopted to to automate the selection of parameters of vehicular shards. Lu et al.~\cite{DBLP:journals/tvt/LuHZMZ20} propose an asynchronous federated learning scheme by adopting DRL for node selection to improve the efficiency.

\section{Blockchain-empowered Heterogeneous Multi-Aggregator Federated Learning Architecture}
\label{sec:framework}
In this section, we describe the proposed \textbf{Blockchain-empowered Heterogeneous Multi-Aggregator Federated Learning Architecture} (BMA-FL). Comparing to existing common FL architecture in edge computing, BMA-FL incorporates multiple aggregators that are connected with different numbers of trainers, and avoids using trainers (edge devices) to maintain the blockchain as trainers are considered have limited resources for maintaining blockchain and train the model at the same time. 

\begin{figure}[htb]
    \centering
    \includegraphics[width = 0.8\linewidth]{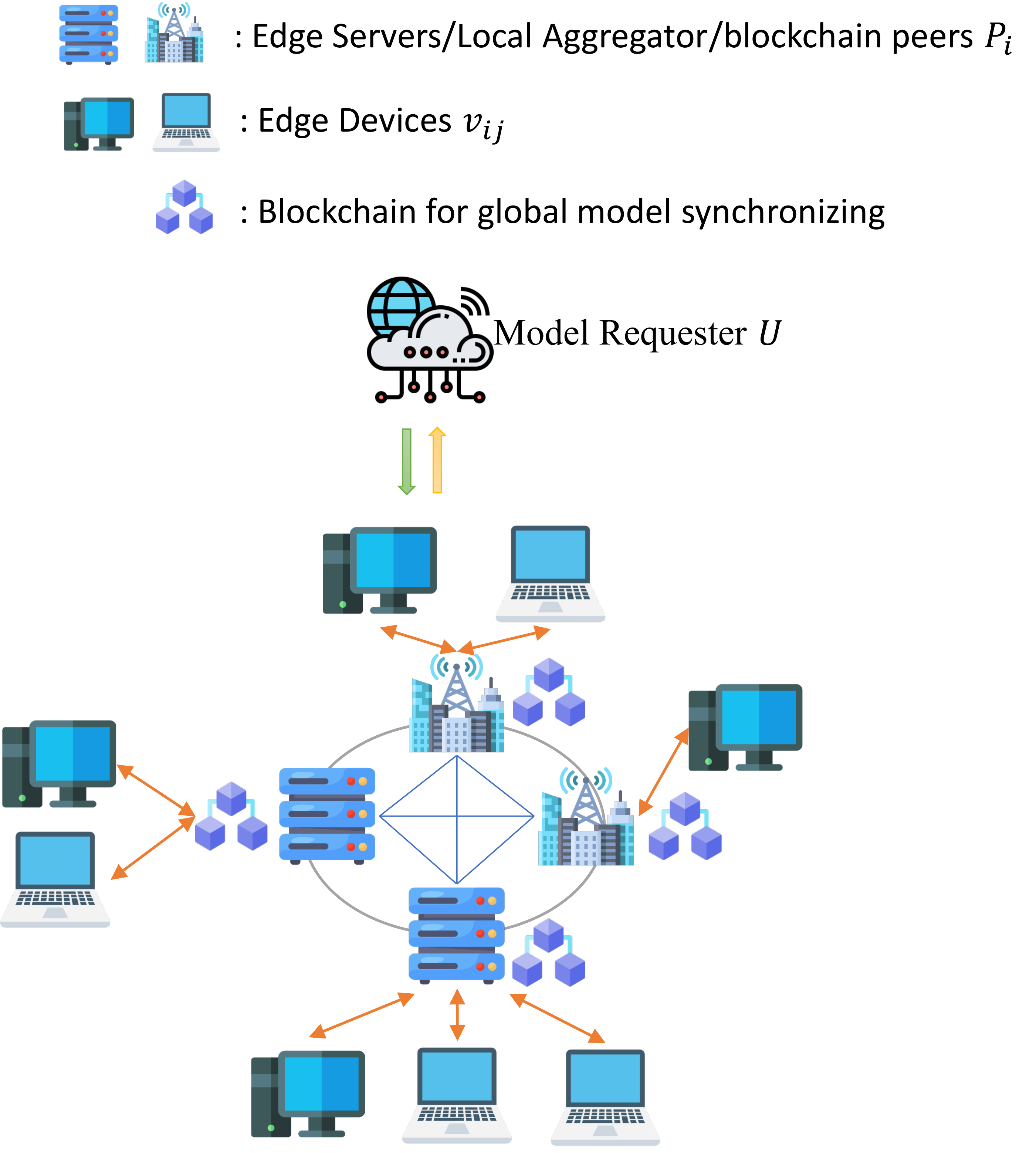}
    \caption{Blockchain-empowered Heterogeneous Multi-Aggregator Federated Learning Architecture (BMA-FL)}
    \label{fig:FLML}
\end{figure}

Figure~\ref{fig:FLML} shows the proposed BMA-FL architecture. In the architecture, there are 3 types of actors: Model Requester, Local Aggregators and Trainers. 

\begin{itemize}
    \item Model Requester: A third-party who distributes a machine learning task. 
    \item Local Aggregators: Edge servers who will receive the task from model requester, and distribute the task to its connected trainers. The test dataset $D$ received from model requester will not be sent to workers. Edge servers are also Peers/Miners for maintain the blockchain. The blockchain is crucial for updating the global model and organizing the model synchronizing. Blockchain module detail will be presented in next section.  
    \item Trainers: The Edge devices who holds training data. They will receive the task from edge server. Trainers train the model parameters $\bm{\theta}$ on their local data and return the trained model parameters back to connected edge servers for local aggregation. 
\end{itemize}

Let $U$ be a model requester that announces a machine learning task that requires the distributed learning framework to train on their owned data.  Let $P_i$ ($ i \in \mathbb{Z^{+}}_N$) be a edge server that connected with multiple trainers $v_{ij}$ ($ j \in \mathbb{Z^{+}}_{m_i}$).

The trainers hold the data, that are not feasible to be massively transmitted or disclosed, therefore the model must be trained locally. The trainers in edge computing context are usually edge devices or IoT micro-computers that have limited computation resources. Edge devices' main function is to collect data and conduct lightweight computing tasks,  therefore those devices should not take the load of maintaining the blockchain system. This issue is often ignored by existing literature. 

Furthermore, much of the current literature presupposes that trainers connect exclusively to a single aggregator (referred to as an edge server in this paper). Such an assumption can be an oversimplification when compared to real-world IoT applications. In these applications, sensors or micro-computers are typically dispersed over a vast area, often connecting to multiple edge servers for communication. In this paper, we consider that multiple edge servers who are connected to multiple edge devices and each edge device will only connect to one edge server. 

BMA-FL executes a machine learning task as following steps in an iterative manner. 
\begin{enumerate}
    \item First, the model requester $U$ will announce the new task $T$ to all edge servers. 
    \item Then, edge servers decide if the connected trainers have satisfied requirements for training the model according to the specifications in $T$. Edge server will hold the test data $D$, and distribute the task to connected feasible trainers. 
    \item Upon receiving the task, trainers will start train the model on its local data. 
    \item Next, for each edge server $P_i$, it can decide a frequency $f_i$ at which it collects the trained parameters from connected edge devices. 
    It's important to note that $P_i$ might gather parameters from $v_{ij}$ that haven't yet converged. Whenever $P_i$ opts to collect, the trainer is obligated to report its current parameters to its associated local aggregator $P_i$. As highlighted in the Introduction section, excessive training on a local device can lead to local optima in parameters, which may not be beneficial for the overall model performance. Thus, $P_i$ must judiciously determine the frequency $f_i$ to optimize performance. This challenge is addressed using a deep reinforcement learning method discussed in Section~\ref{sec:drl}.
    
    \item After an edge server $P_i$ collects the reported parameters from $v_{ij}$, denoted as $\bm{\theta}_{ij}$, $P_i$ will conduct \textbf{\emph{Local Aggregation}}, which aggregates $\bm{\theta}_{ij}$ into one local model represented by parameter $\bm{\theta}_{i}$. Then $P_i$ distribute this aggregated model back to the connected edge devices $v_{ij}$. Please note that as $f_i$ varies among different $P_i$, hence different $P_i$ may conduct different rounds of local aggregation, which will impact the final trained model. 
    \item Each edge server maintains the local parameters update logs which are stored as transactions, and will be packed into blocks at frequency $F$ which is a given constant. Actually, at a frequency $F$, edge servers need to make consensus on current global model which is aggregated from all local models $\bm{\theta}_{i}$. This step is named \textbf{\emph{Global Aggregation}}. 
    \item The edge server who conduct the global aggregation is chosen through the blockchain module, the global update will  also be propagated and verified through blockchain. 
    \item Upon receiving the new block that contains the updated global model parameters, each edge server will distribute these updated parameters to its connected trainers. Then the trainers will conduct another round of training. Above steps will be repeated until the global model converges or the performance is satisfied. 
\end{enumerate}

\subsection{Blockchain Module Architecture}
\begin{figure*}[htb]
    \centering
    \includegraphics[width = 0.75\linewidth]{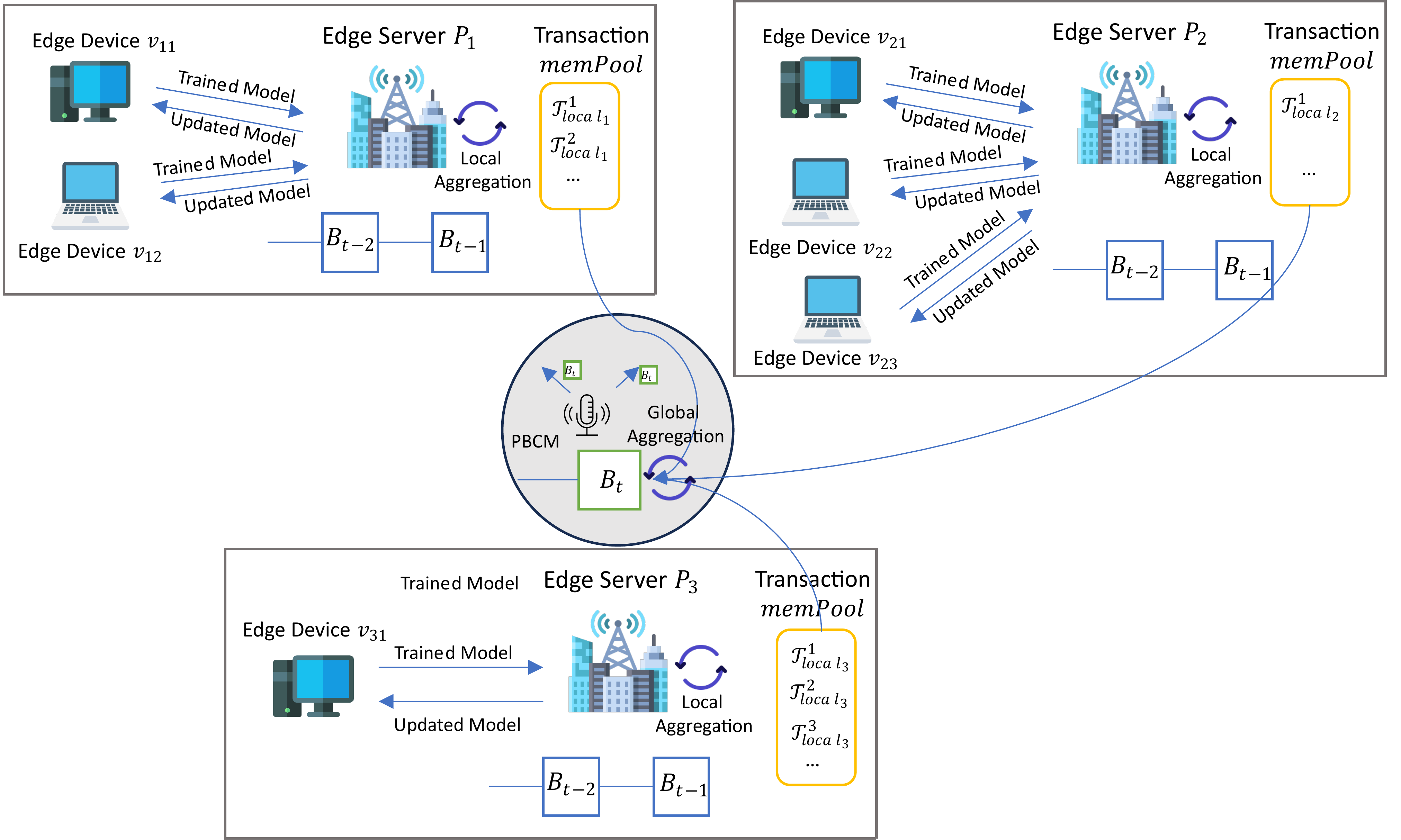}
    \caption{The blockchain module in Blockchain-empowered Heterogeneous Multi-Aggregator Federated Learning Architecture (example of 3 edge servers)}
    \label{fig:blockchain} 
\end{figure*}

The blockchain module in the  BMA-FL architecture is responsible model update logs, conducting global aggregation and transmitting the aggregated global model.
The blockchain module is illustrated in Figure~\ref{fig:blockchain}. 

\begin{figure}[htb]
    \centering
    \includegraphics[width = 0.75\linewidth]{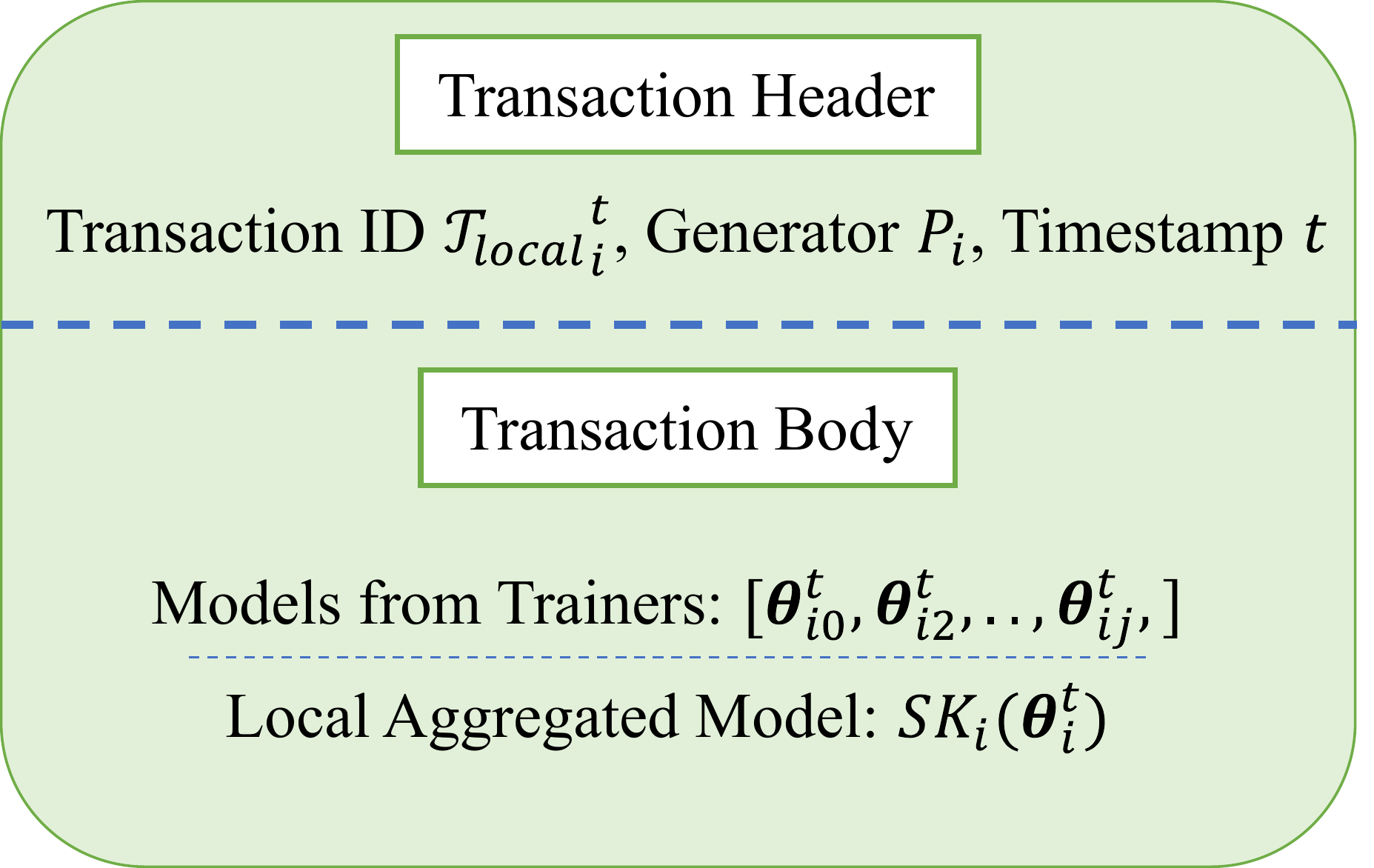}
    \caption{The structure of local aggregation transaction $\mathcal{T}_{{local}_i}$ at timestamp $t$}
    \label{fig:tran}
\end{figure}

As described above, for a local aggregation round, edge server $P_i$ will collect the current parameters $\bm{\theta}_{ij}$ from all connected edge devices $v_{ij}$, and distribute the aggregated model back to edge devices. The updated parameters will be stored in a log, called \emph{Local Aggregation Transaction}. As shown in Figure~\ref{fig:tran}, a \emph{Local Aggregation Transaction} contains timestamp information and parameter records. 

At the frequency $F$ (which means every $F$ units of time, one block will be generated), global aggregation is conducted that a chosen edge server $P_j$ will act as a miner. Let use $k$ to denote the current global aggregation time. The miner will first collect all $\mathcal{T}_{{local}_i}$ from all $P_i$ ($i \neq j$) which obviously contains the latest local model parameters $\bm{\theta}_{i}$, then conduct conventional federated learning aggregation, i.e. average aggregation. $P_j$ will not wait for edge servers finish their on-going training round, but grab all up-to-date $\mathcal{T}_{{local}_i}$. The aggregated model is the global model denoted as $\bm{\theta}^k$ and package them into a \emph{Global Aggregation Block}, denoted as $\mathcal{B}$. For brevity, we use \emph{Block} to refer to \emph{Global Aggregation Block}.

\begin{figure}[htb]
    \centering
    \includegraphics[width = 0.99\linewidth]{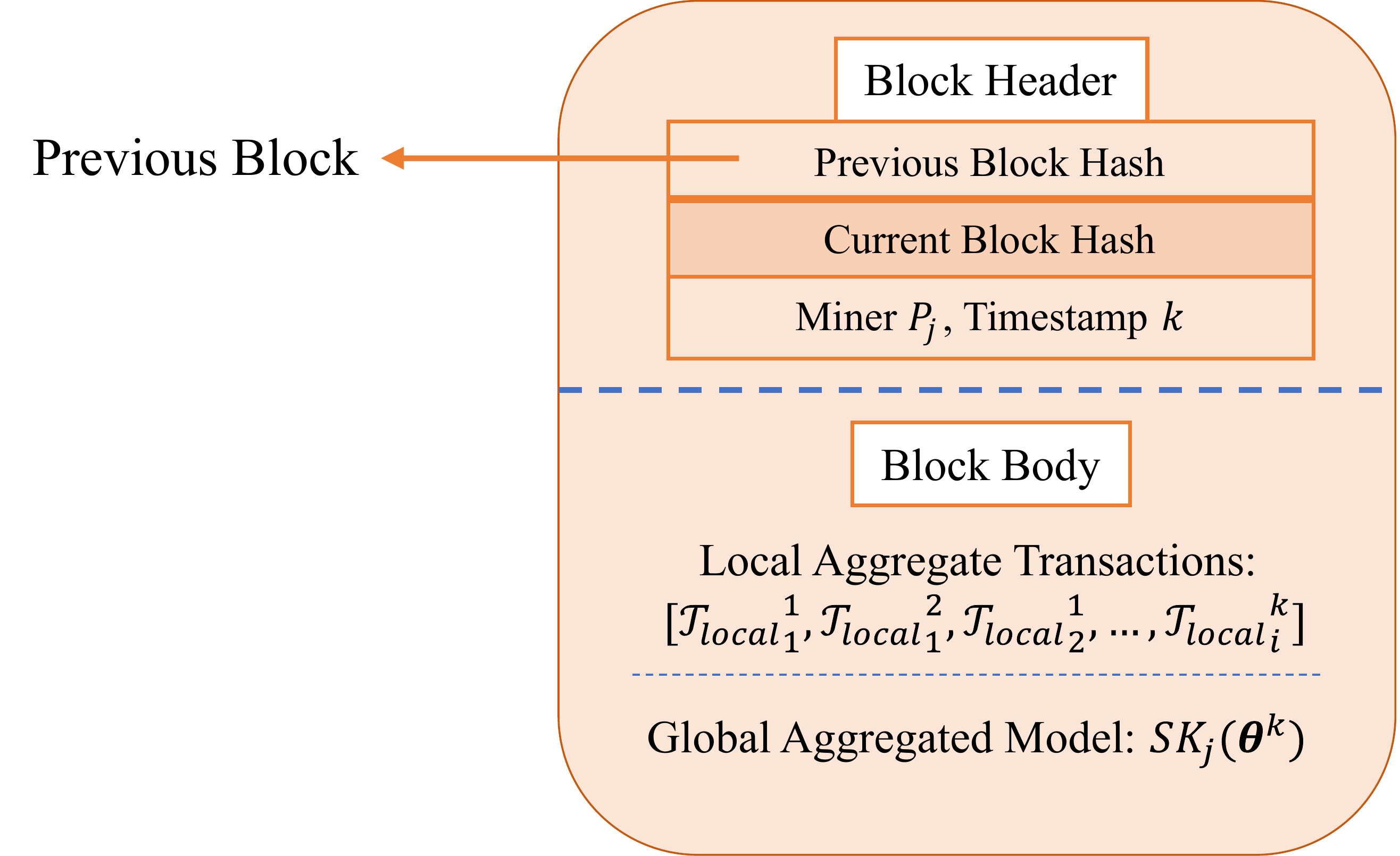}
    \caption{The block storage structure of $\mathcal{B}$ at timestamp $k$}
    \label{fig:block}
\end{figure}

The structure of  $\mathcal{B}$ is shown in Figure~\ref{fig:block}. $\mathcal{B}$ contains block header and block body. Block header records the miner's identity information, timestamp and previous block's hash value. The block body records all $\mathcal{T}_{{local}_i}$ for $ i \in \mathbb{Z^{+}}_N$ and the aggregated global model  $\bm{\theta}^k$. The miner edge server $P_j$ will then broadcast the created block for all other edge servers following designed consensus mechanism proposed in next section. 

\subsection{Performance-based Byzantine Consensus Mechanism}
The consensus mechanism in the blockchain module is crucial for choosing trustable edge servers to aggregate models and achieve consensus of global models among all edge servers. 

Compared with traditional federated learning architecture where the situation is straightforward that the only aggregator will aggregate the collected models and derive a global model, in multi-aggregators setting, we have to design a consensus among all aggregators to elect one aggregator at each global aggregation round to do model aggregation as well as to make it synchronized and verified by all other aggregators. 

A common consensus mechanism is Proof-of-Work (PoW). However it will waste considerable energy on solving meaningless hash puzzle. As raised by~\cite{DBLP:journals/tpds/QuWHC21}, the energy consumed on hash puzzle can be saved for executing federated learning tasks. On the other hand, due to lack of trust and delay or disturb of communication among edge servers, a byzantine fault tolerant consensus is desired in this paper's scenario.  Castro and Liskov~\cite{castro1999practical} proposed a practical byzantine fault tolerance algorithm (PBFT) that can achieve final consensus after three specific message propagation. PBFT is now commonly used in blockchain systems~\cite{DBLP:journals/tgcn/GuoDW22}. However, PBFT does not includes the miner election as required in our proposed architecture. In addition, a corresponding incentive mechanism should exist in the consensus mechanism to motivate the edge servers to work hard for providing helpful local models that benefit the global model. 

\begin{figure}[htb]
    \centering
    \includegraphics[width = 0.99\linewidth]{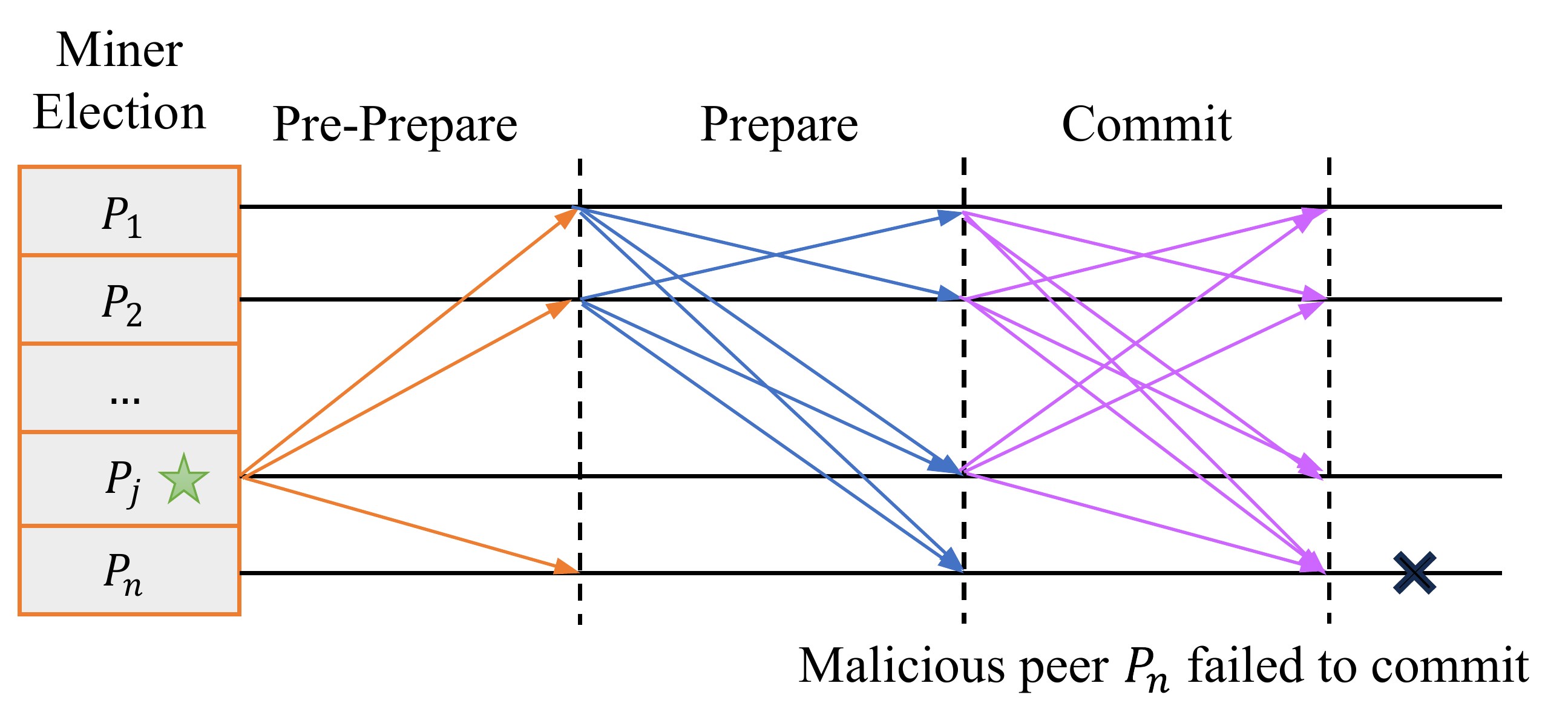}
    \caption{Illustration of PBCM}
    \label{fig:PBCM}
\end{figure}

To solve above problems, this paper proposes \textbf{\emph{Performance-based Byzantine Consensus Mechanism}} (PBCM). PBCM contains 4 stages: miner election, pre-prepare, prepare and commit as illustrated in Figure~\ref{fig:PBCM}.

\emph{(1): Miner Election}: One edge server will be elected based on their model training performance as the miner. We evaluate the training performance of an edge server by the amount of its performance increase ($PI$) in terms of evaluation metrics specified as $E$ in learning task $T$, e.g. model accuracy. Specifically $PI_i^k = E(\bm{\theta}_{i}^{k}) - E(\bm{\theta}_{i}^{k-1})$, where $\bm{\theta}_{i}^{k}$ is the current up-to-date aggregated local model of $P_i$ at $k$-th global aggregation round and $\bm{\theta}_{i}^{k-1}$ is the previous local model of $P_i$ at $(k-1)$-th global aggregation round. 

We then define a trust score $S_i^k$ associated with $P_i$. Based on the common sense that edge servers who train better models are more trust-worthy and who honestly conduct mining and verification in history are more trust-worthy, the $S_i^k$ is defined as Equation~\ref{eq:sik1} and Equation~\ref{eq:sik2}. The trust score is updated after each round of global aggregation. Based on the different role of $P_i$ at round $k$, the trust score is updated accordingly. 

When $P_i$ is the miner at round $k$: 


\begin{equation}
\label{eq:sik1}
    S_i^{k} = 
\begin{cases} 
 \min\{ 1, S_i^{k-1} +\Delta_1\} + PI_i^k & \mbox{if } \mathcal{B}^k \mbox{ is accepted} \\
\max\{ 0, S_i^{k-1} -\Delta_1\} + PI_i^k  & \mbox{if } \mathcal{B}^k \mbox{ is rejected} .
\end{cases}
\end{equation}    	

When $P_i$ is no the miner at round $k$:
\begin{equation}
\label{eq:sik2}
    S_i^{k} = 
\begin{cases} 
 \min\{ 1, S_i^{k-1} +\Delta_2\} + PI_i^k & \\ \mbox{ if } P_i \mbox{'s decision consistant with final result} \\
\max\{ 0, S_i^{k-1} -\Delta_2\} + PI_i^k  & \\ \mbox{ if } P_i \mbox{'s decision incosistant with final result} .
\end{cases}
\end{equation}    	

We set $\Delta_1>\Delta_2>0$, as the rewards to miner should be higher than other peers in blockchain because most times edge devices are just regular peers. The initial parameters $PI_i^0 = 0$ and $S_i^{0}=0$. At the $k$-th global aggregation round, the miner is selected from all edge servers $P_i$ ($i \in \mathbb{Z^{+}}_N$) that an edge server $P_i$ is selected with probability ${Pr}_i^k$:
\begin{equation}
    {Pr}_i^k = \frac{S_i^{k-1}}{\sum_{j\in \mathbb{Z^{+}}_N} S_j^{k-1}}
\end{equation}
Obviously, the edger sever who has better training effects (higher performance increase) will have higher probability to be selected as miner. A miner or a regular peer who verifies the block will also gain rewards if it provides valid block or block verification, consequently the trust score will increase. 

\emph{(2): Pre-prepare}: At this stage, the selected miner $P_j$ will package the block $\mathcal{B}^k$, and create a pre-prepare message $(SK_j(\mathcal{B}^k),pre-prepare)$. $SK_j$ is the secrete key of $P_j$. $P_j$ then broadcasts $\mathcal{B}^k$ and the pre-prepare message to all other edge servers and requests responses. All these edge servers will verify if the received block $\mathcal{B}^k$ contains valid information, e.g. the parameters in the transaction are in valid format, no anomaly in the history of parameter updates. If the block is verified valid, it will be stored, otherwise it will be discarded.

\emph{(3): Prepare}: If a non-miner edge server $P_i$ believes  $\mathcal{B}^k$ is valid, it will broadcast a prepare message $(SK_i(SK_j(\mathcal{B}^k)),prepare)$ to all the other edge servers, so that everyone knows  $P_i$'s decision.  Then every edge server (including the miner) gathers all the received prepare message. All edge servers must make initial decisions at this step, that whether add or discard this block $\mathcal{B}^k$ into existing blockchain. Let $G_i^k$ be edge server set from which edge server $P_i$ receives prepare messages (including $P_i$ itself), we let $P_i$ accept  $\mathcal{B}^k$  if following statement stands:
\begin{equation}
    \sum_{n\in G_i^k} {Pr}_n^k \geq (2\lfloor \frac{N-1}{3} \rfloor +1)\frac{1}{N},
\end{equation}
where $N$ is the total number of edge servers in the system. If $P_i$ accepts  $\mathcal{B}^k$, it will broadcast commit message $(SK_i(SK_j(\mathcal{B}^k)),commit)$ to all other edge servers. 

\emph{(4): Commit}: After edge servers send out commit message, they will wait commit messages from other edge servers,. For each edge server $P_i$, its consensus process is called completed when its received commit messages satisfies the following statement: 
\begin{equation}
    \sum_{n\in H_i^k} {Pr}_n^k \geq (2\lfloor \frac{N-1}{3} \rfloor +1)\frac{1}{N},
\end{equation}
where $H_i^k$ is edge server set from which edge server $P_i$ receives commit messages (including $P_i$ itself). At this point, $P_i$ knows the new block $\mathcal{B}^k$ is accepted by the whole blockchain module and will append the new block into its local blockchain no matter what was the $P_i$'s intial decision. The up-to-date model parameters as well as the update history will be permanently, unalterably stored in the blockchain shared with all edge servers, which provides transparency and security to the BMA-FL architecture.

After the commitment, all edge servers $P_i$ will update their score $S_i^{k}$ according to Equation~\ref{eq:sik1} and Equation~\ref{eq:sik2} for next round miner election and blockchain update.

\section{Deep Reinforcement Learning Optimization Scheme}
\label{sec:drl}

To be chosen as miners, edge servers are incentivized to enhance their training outcomes through local aggregation. Within the distributed learning framework, the Non-IID data distribution and CPU speed vary among trainers, leading to a natural variation in model quality. However, classic FL architectures often weight the model based on the data volume of each trainer, neglecting the actual quality of the training~\cite{DBLP:conf/aistats/McMahanMRHA17}. Furthermore, these architectures tend to wait for the slowest trainer to complete its training before proceeding with aggregation. We address this practical issue in this section, that the model aggregation should not only take trainers' data amount as weight, but also take account of the performance of their actual trained model.

In addition, blockchain module appends new block with a given frequency $F$, local aggregators have to trade off how many times of local aggregation should they perform, as more local aggregation rounds bring better parameter sharing among trainers but the parameters will be less converged on each trainer as less training time per local aggregation. Let $f_i$ be the local aggregation frequency decided by edge server $P_i$, which means every $f_i$ units of time, $P_i$ will perform one local aggregation. In this paper, we devise an deep reinforcement learning optimization scheme that selects the most appropriate weights for trainers during local aggregation, aiming to expedite model training.


For a trainer $v_{ij}$, it holds data denoted as $D_{ij}$, it have CPU speed for training data $c_{ij}\ epoch/unit >0$ that is the $v_{ij}$ can train $c_{ij}$ epochs on data $D_{ij}$ within each time unit. For example, $c_{ij}=0.5$ means it $v_{ij}$ can only finish half of the epoch given data $D_{ij}$. 

Though in FL, the data are considered privacy and should not be disclosed to anyone, in this paper, we give the right back to edge devices and edge servers, so that edge devices can decide if to upload the data to edger servers for getting more data trained, or keep the data private locally.  To enable modeling different use cases, we set up the penalty of data upload as $\sigma*|D_{{up}_{ij}}|$. Where $\sigma$ is the penalty coefficient and $|D_{{up}_{ij}}|$ is the size of uploaded data from $v_{ij}$ to $P_i$. In this paper, we only allow $D_{{up}_{ij}}$ be the portion of data that $v_{ij}$ not able to finish within one epoch during $f_i$. The uploaded data will not be trained again locally at $v_{ij}$. $\sigma$ can be adjusted based on real application demands, higher $\sigma$ will lead the system try to avoid any data upload, and lower one can allow more data offloaded. Edger server $P_i$ will decide if the uploaded data $|D_{{up}_{ij}}|$ will be accepted based on $v_{ij}$'s training performance because if  $v_{ij}$ can produce good models, the data from  $v_{ij}$ are more likely considered valuable. Specifically, let $\bm{\theta}_{ij}$ be the trained model of an edge device $v_{ij}$, then a decision factor $U_{ij}$ is calculated by $P_i$ to determine the acceptance. 
\begin{equation}
    U_{ij} = \sigma*|D_{{up}_{ij}}| + h_{i1}*E(\bm{\theta}_{ij}) + a_i.
\end{equation}
 $h_{i1}$ and $a_i$ are parameters to be decided by $P_i$. If $U_{ij}>=0$,  $P_i$ will accept the uploaded $|D_{{up}_{ij}}|$ and train it on edge server. Otherwise, $P_i$ will refuse to train $|D_{{up}_{ij}}|$. 

At each local aggregation, edge server $P_i$ assigns weight $W_{ij}$ to $\bm{\theta}_{ij}$ for local aggregation. We first define $W_{ij}'$ to combine the impact of data amount and the trainers model accuracy as:

\begin{equation}
    W_{ij}' = w_{i0}*|D_{ij} \setminus D_{{up}_{ij}}| + w_{i1}*E(\bm{\theta}_{ij}) + b_i,
\end{equation}
where $w_{i0}$, $w_{i1}$ and $b_i$ are parameters to be decided by $P_i$.  Then $W_{ij}$ is further defined in Equation~\ref{eq:W} as normalized $ W_{ij}'$ so that $sum_{j \in \mathbb{Z^{+}}_{m_i}} W_{ij} = 1$. 

\begin{equation}
\label{eq:W}
    W_{ij} = \frac{W_{ij}'}{\sum_{j \in \mathbb{Z^{+}}_{m_i}} W_{ij}'}
\end{equation}

Finally the local aggregation is defined in Equation~\ref{eq:locl_agg}, that the aggregated model $\bm{\theta}_{i}$ at $P_i$ is the averaged trainer models $\bm{\theta}_{ij}$ weighed by $W_{ij}$.
\begin{equation}
\label{eq:locl_agg}
    \bm{\theta}_{i} = \sum_{j \in \mathbb{Z^{+}}_{m_i}} W_{ij}*\bm{\theta}_{ij}
\end{equation}

$W$ allows edge servers dynamically adjust the weights to the collected models from edge devices based on each rounds model performance, which will motivate each trainers keep producing effective models. Consequently, the final model is more robust to be spoiled by malicious trainers, and the whole FL schema is more secure. 

To summarize the objective of edge server $P_i$, $P_i$ has to decide six key parameters, including $f_i$, $h_{i1}$, $a_i$, $w_{i0}$, $w_{i1}$ and $b_i$ so that $P_i$ is able to produce better aggregated model during each block interval $F$, which will help $P_i$ earn more probability to become miner at blockchain consensus phase. 

\begin{figure*}[htb!]
    \centering
    \includegraphics[width = 0.6\linewidth]{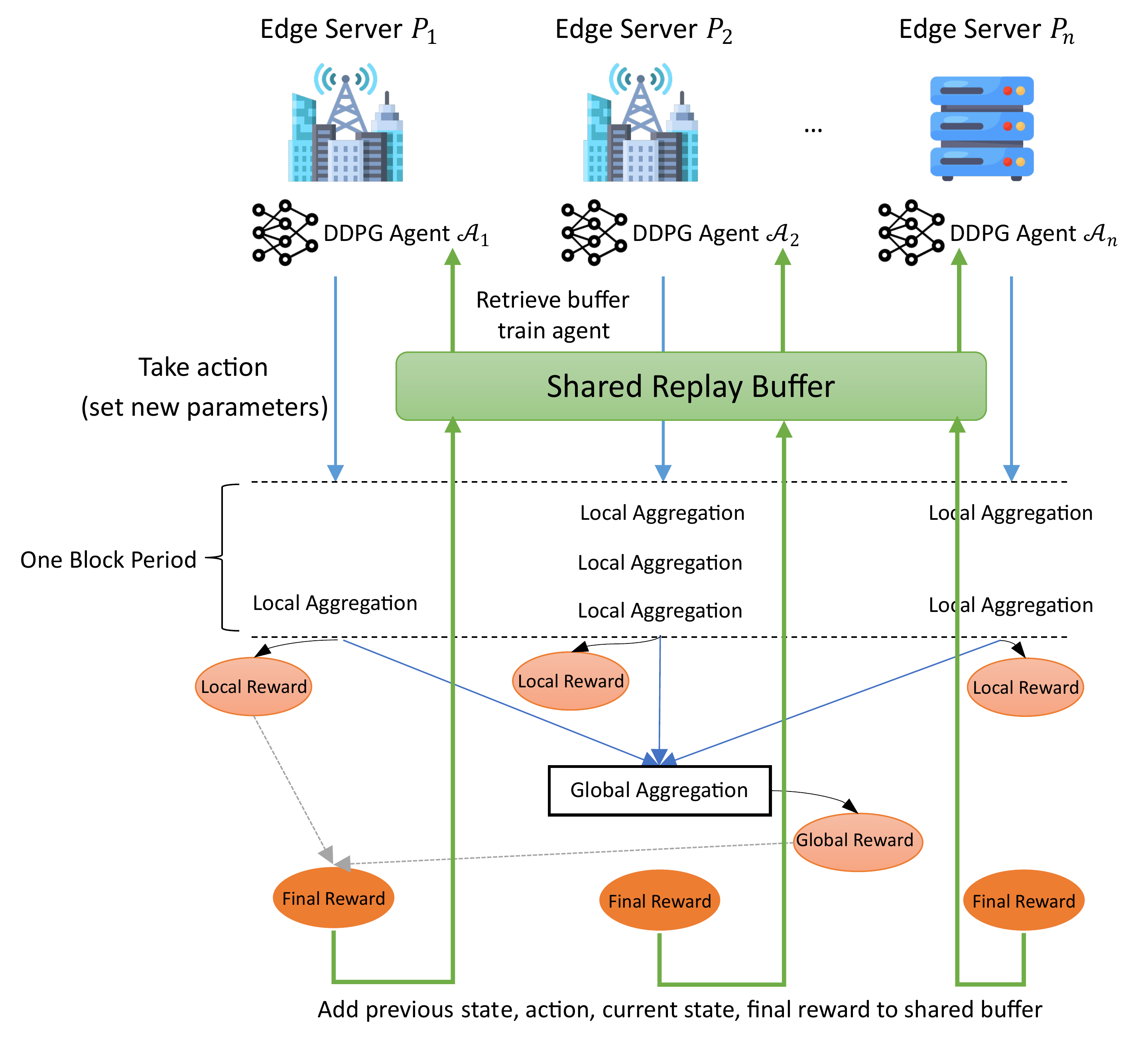}
    \caption{Multi-Agent with Shared Buffer Deep Reinforcement Learning algorithm }
    \label{fig:DDPG_sb}
\end{figure*}

In this paper, we design a \textbf{Multi-Agent with Shared Buffer Deep Reinforcement Learning algorithm }(MASB-DRL) to solve the problem. The general idea is that for each edge server $P_i$, a reinforcement learning (RL) task is defined and will be solved by a separate RL agent $\mathcal{A}_i$ that the $\mathcal{A}_i$ is able to learn from not only $P_i$'s past experience but from all other edge servers through \emph{Shared Replay Buffer}. The architecture of MASB-DRL is illustrated in Figure~\ref{fig:DDPG_sb}. The challenge of the RL task is how to make the past experience from different edge servers useful for others as they are within different RL environment settings, i.e. varied number of connected edge devices, with varied data amount and training speed on each. The same parameter setting for different edge server may result in totally different outcomes. MASB-DRL addresses this challenge through incorporating the edge servers status into agent training status. 

In detail, the proposed MASB-DRL on each edge server $P_i$ is defined as follows. 

\textbf{Agent} $\mathcal{A}_i$: Each edge server $P_i$ has an agent $\mathcal{A}_i$. Because our action space is not discrete,  we adopt Deep Deterministic Policy Gradient (DDPG) algorithm~\cite{DBLP:journals/corr/LillicrapHPHETS15} as agent, which is a popular and state-of-art deep RL algorithm for continuous action space.

\textbf{Action Space} $A_i$: $A_i$ is composed of the six parameters, $A_i = [f_i, h_{i1}, a_i, w_{i0}, w_{i1}, b_i]$, agent $\mathcal{A}_i$ is able to take continuous action for each parameter such that $f_i\in (0,F]$, $h_{i1} \in (0,1]$, $a_i \in [-1,1]$, $w_{i0}\in (0,1]$, $w_{i1}\in(0,1]$ and $b_i \in [-1,1]$. 

\textbf{State Space} $St_i$: $St_i = [ E(\bm{\theta}^{k}),  E(\bm{\theta}_{i}^{k}), |D_i|, N]$, where $D_i = \sum_{j \in \mathbb{Z^{+}}_{m_i}} D_{ij}$ and $N$ is the total number of edge devices connected to $P_i$. $E(\bm{\theta}^{k})$ and $ E(\bm{\theta}_{i}^{k})$ are the current evaluation of learnt global model and local model, respectively. Please note $E(\bm{\theta}^{k})$ is calculated after the global aggregation. It is important to highlight that $|D_i|$ and $N$ are actually constant. Normally, constants in state in RL are meaningless, however, $|D_i|$ and $N$ are crucial here to describe the edge server, so that other edge servers with similar setting can use the experience of $\mathcal{A}_i$, e.g. (state, action) pairs. Future search can propose more sophisticated descriptive variables as substitutions of $|D_i|$ and $N$. 

\textbf{Reward} $R$: The reward is to evaluate how the actions $A_i$ taken affect both its local model and the final global model. Therefore, $R$ is calculated after the global aggregation when we evaluate the aggregated global model. 
Let $E(\bm{\theta}^{k-1})$ be the initial local model performance before it is trained by $P_i$ at this global aggregation round $k$. We define a local reward $R_{l}$ as Equation~\ref{eq:r_l}, to denote the immediate reward for gaining improvement of local model. 
\begin{equation}
\label{eq:r_l}
    R_{l} = p*[E(\bm{\theta}_{i}^{k}) - E(\bm{\theta}^{k-1})]*E(\bm{\theta}_{i}^{k})
\end{equation}

$p$ is a constant coefficient, $E(\bm{\theta}_{i}^{k}) - E(\bm{\theta}^{k-1})$ reflects how much the local model on $P_i$ is improved during this round of global aggregation given action  $A_i$. Considering the model will become harder to improve when it reaches to a relatively high performance, $*E(\bm{\theta}_{i}^{k})$ helps increase the reward when model reaches high performance. 

Similarly, we define a global reward $R_g$ to denote the impact of $A_i$ on the improvement of global model as in Equation~\ref{eq:r_g}. 
\begin{equation}
\label{eq:r_g}
    R_{g} = q*[E(\bm{\theta}^{k}) - E(\bm{\theta}^{k-1}))]*E(\bm{\theta}^{k})
\end{equation}

$q$ is a constant coefficient and $E(\bm{\theta}_{i}^{k}) - E(\bm{\theta}_{i}^{k-1})$ reflects how much the local model on $P_i$ is improved during this round of global aggregation given action  $A_i$.

Finally, the total reward at this step is defined in Equation as a product of $R_l$ and $R_g$. 

\begin{equation}
\label{eq:r}
    R = R_{l}*R_{g}
\end{equation}

\begin{figure}[htb]
    \centering
    \includegraphics[width = 1\linewidth]{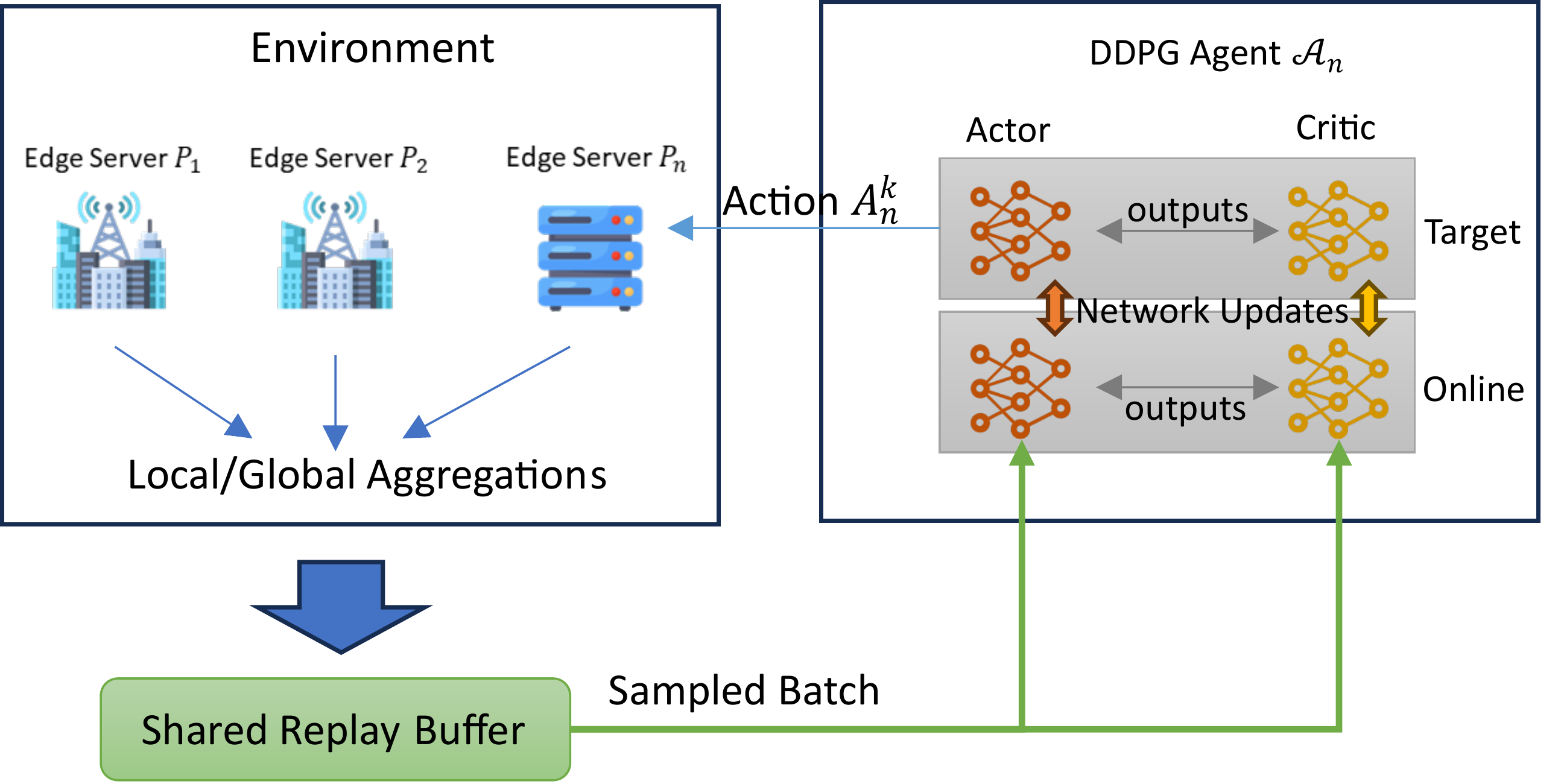}
    \caption{Illustration of DDPG Algorithm}
    \label{fig:ddpg}
\end{figure}

\textbf{Shared Replay Buffer} $B$: DDPG Agent $\mathcal{A}_i$ is able to learn from past execution history,  e.g. replay buffer, with its critic neural network and actor neural network. The shared repaly buffer $B = \{ [St_i^k, A_i^k, R^k,St_i^{k+1}]| k\in \mathbb{Z}  \}$. The detail about how DDPG agents learn from replay buffer is illustrated in Figure~\ref{fig:ddpg}.

\section{Experiments}
\label{sec:eval}

In this section, we evaluate the performance of proposed Blockchain-empowered Heterogeneous Multi-Aggregator Federated Learning Architecture on 3 common real-world datasets. We then evaluate the effectiveness of Multi-Agent with Shared Buffer Deep Reinforcement Learning algorithm (MASB-DRL) by comparing it with two baselines. We finally evaluate the robustness of Performance-based Byzantine Consensus Mechanism (PBCM). 

\subsection{Baselines}

\begin{itemize}
    \item Central: Traditional centralized training mode, all data are located on a powerful server and trained on single machine. 
    \item FedAvg~\cite{DBLP:conf/aistats/McMahanMRHA17}: Uses the proposed BMA-FL architecture, but all parameters are fixed in all rounds. i.e. $A_i = [f_i, h_{i1}, a_i, w_{i0}, w_{i1}, b_i]=[1,1,1,1,1,1], \forall i\in N$. Both local aggregation and global aggregation use FedAvg algorithm. 
    \item Random: Uses the same proposed BMA-FL architecture, but after first round, parameters in $A_i,\forall i\in N$, are randomly picked within the value boundaries. 
    \item BMA-FL: Uses BMA-FL architecture and our proposed MASB-DRL method. Deep reinforcement learning agents will learn the best parameters based on shared experience after each round and take the best actions in next round.
\end{itemize}
In the following sections, all results are reported as the average of 10 repeated executions. 

\subsection{Datasets}
We use three datasets: MNIST~\cite{lecun2010mnist}, FashionMNIST~\cite{DBLP:journals/corr/abs-1708-07747} and CIFAR-10~\cite{Krizhevsky09learningmultiple} which are popular in machine learning and federated learning research. We conduct image classification tasks on the datasets as our federated learning task for evaluation. 

\begin{itemize}
    \item MNIST: Contains a training set of 60,000 image examples and a test set of 10,000 image examples of handwritten digits (from 0-9). Each example is a 28x28 grayscale image. 
    \item FashionMNIST: Contains of a training set of 60,000 examples and a test set of 10,000 examples of different cloth styles (e.g. T-shirt, Trouser, etc.). Each example is a 28x28 grayscale image, associated with a label from 10 classes. 
    \item CIFAR-10: Contains a training set of 50,000 image exmaples and a test set of 10,000 image examples of different objects (e.g. Airplane, Cat, Ship, etc.). Each example is a 32x32 color image, associated with a label from 10 classes.
\end{itemize}

We choose the above three datasets to simulate different task difficulty levels as easy, medium and hard, respectively. For MNIST and FashionMNIST we aim to train a neural network as containing two convolutions layer, two dropout layer and two linear layer as suggested in Pytorch's official manual\footnote{https://github.com/pytorch/examples/tree/main/mnist}. For CIFAR-10, we aim to train a ResNet9 model\footnote{https://github.com/davidcpage/cifar10-fast}.

\subsection{Experiments Settings}
We implemented a Blockchain-empowered Heterogeneous Multi-Aggregator Federated Learning Architecture (BMA-FL) with one model requester, five edge servers as local aggregators. The five edge servers are connected with 2, 4 ,6, 8 and 10 edge devices respectively. The CPU speed $c_{ij}$ of each edge device are also varied, and is set $(1.0, 0.5)$, $(2.5, 2.0, 3.0, 2.0)$, $(2.5, 3.0, 0.5, 0.5, 3.0, 3.0)$, $(3.0, 2.5, 2.0, 2.5, 3.5, 3.5, 3.0, 3.5)$, $(2.0, 3.5, 2.0, 1.0, 0.5, 2.0, 3.5, 1.0, 3.5, 3.5)$ respectively. The global aggregation frequency $F = 2$.

All the training sets from three datasets are manually divided unevenly and in heterogeneous label distribution (Non-IID) on each edge device. The test sets are treated as $D$, which is from model requester $U$ for evaluating the final performance of trained model. In PBCM, $\Delta_1 = 2$ and $\Delta_2 = 1$. The offloading penalty $\delta = 4$.

\subsection{Convergence Evaluation}
\begin{figure}[htb]
    \centering
    \includegraphics[width = 0.8\linewidth]{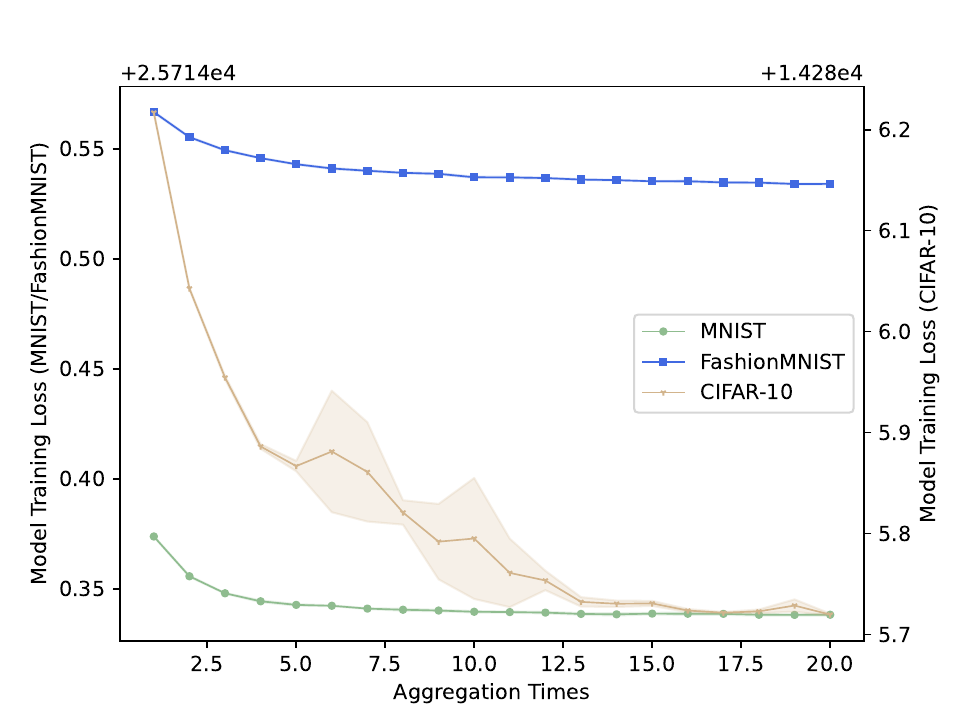}
    \caption{The trained model loss after each global aggregation in BMA-FL}
    \label{fig:loss}
\end{figure}

We first evaluate the performance of proposed MBA-FL on training machine learning models. Figure~\ref{fig:loss} shows how the BMA-FL converges on training the three models on above three datasets. 

We use Accuracy as the metric to evaluate the final converged global models. The comparison among baselines are reported in Table~\ref{tab:per}. We can see all federated learning architecture show slightly lower performance than traditional centralized training schema. This is because the gradients can not be accurate propagated to all data samples and the Non-IID data distribution brings challenge on generalizability of trained models. As image classification on CIFAR-10 is the hardest task among all three, the decentralized baselines show the most accuracy gap to the centralized training architecture on CIFAR-10. data  Among all FL architecture, our proposed BMA-FL can achieve the best performance. This demonstrates our proposed deep reinforcement learning algorithm MASB-DRL can effectively learn better training strategies in BMA-FL, aiding the aggregators in BMA-FL better decide local aggregation frequency and balance the weights for the collected models at local aggregation. 

\begin{table}[hbp]
\centering
\caption{The performance proposed FL Architecture }
\label{tab:per}
\begin{tabular}{c|ccc}
\hline
Method   & MNIST & FashionMNIST & CIFAR-10 \\ \hline
Central   &      99.18\%     &   93.05\%     &        93.37\%             \\ \hline
FedAvg  &      99.05\%     &   91.87\%   &       90.50\%      \\
Random  &     99.01\%      &   91.20\%  &        89.88\%        \\
\textbf{BMA-FL}  &    \textbf{99.10\% }     &    \textbf{92.04\% }     &         \textbf{91.70\%}         \\ \hline
\end{tabular}
\end{table}

\begin{figure}
	\centering
	\subfigure[MNIST]{
		\begin{minipage}[b]{0.4\textwidth}
			\includegraphics[width=1\textwidth]{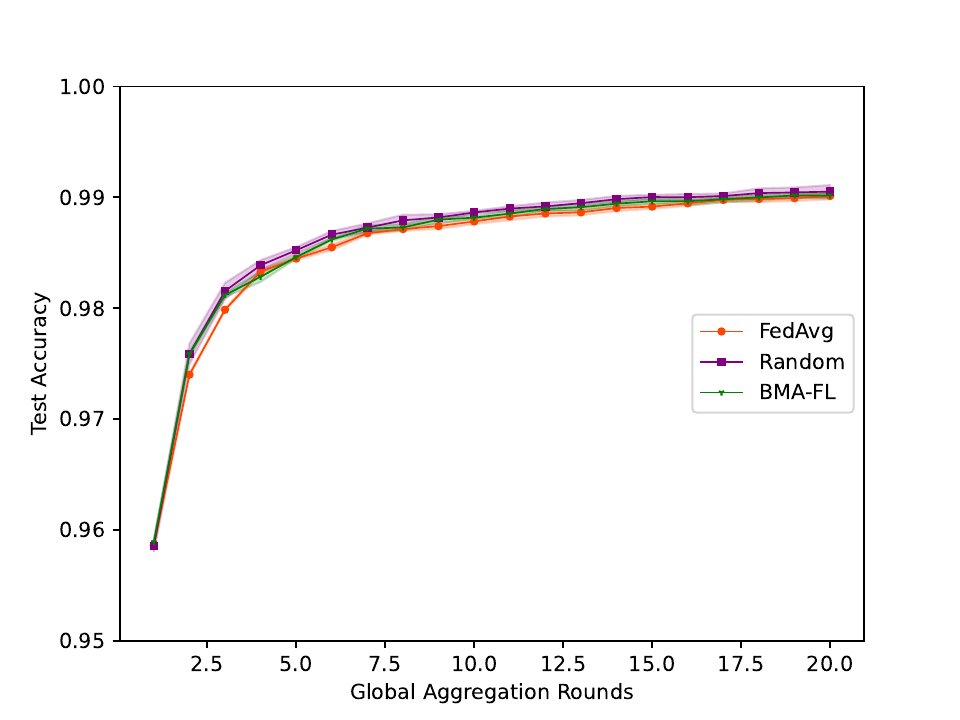}
		\end{minipage}
		\label{fig:speed_mnist}
	}
 \\
    \subfigure[FashionMNIST]{
    		\begin{minipage}[b]{0.4\textwidth}
   		 	\includegraphics[width=1\textwidth]{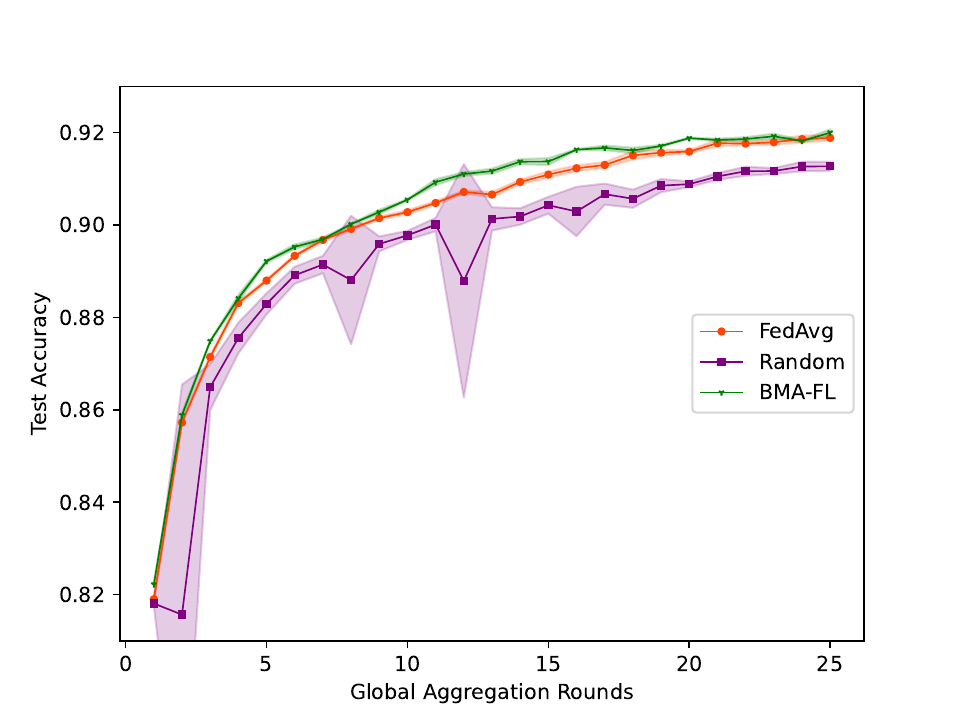}
    		\end{minipage}
		\label{fig:speed_fmnist}
    	}
     \\
     \subfigure[CAFIR-10]{
    		\begin{minipage}[b]{0.4\textwidth}
   		 	\includegraphics[width=1\textwidth]{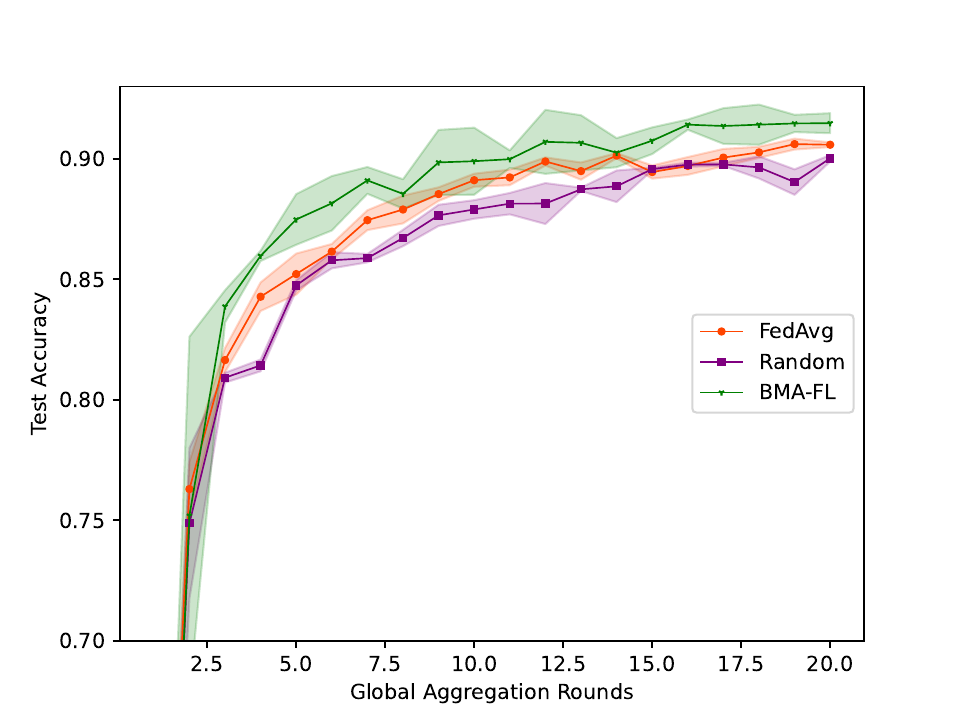}
    		\end{minipage}
		\label{fig:speed_cafir}
    	}
	\caption{Model Convergence Speed Comparison}
	\label{fig:speed}
\end{figure}

\subsection{Performance of MASB-DRL}

Figure~\ref{fig:speed} shows the model accuracy on test datasets after each global aggregation. As the blockchain generating frequency is fixed and each block generation produces single global model consensus, therefore the less global aggregation rounds before convergence means less time spent on training.  

In Figure~\ref{fig:speed_mnist}, three methods show marginal differences. Even randomly choosing the parameters does not have too much negative impact on the result. This is because image classification task on MNIST dataset is easy and not sensitive the parameters.
Figure~\ref{fig:speed_fmnist} and Figure~\ref{fig:speed_cafir} shows more clear difference among three algorithms. It is obvious that our proposed BMA-FL shows faster convergence speed than FedAvg and  ``Random". More specifically, on FashionMNIST dataset, BMA-FL used averagely only 20 global aggregation rounds to reach the performance of FedAvg at 25th global aggregation round, which means BMA-FL is 20\% faster than FedAvg on FashionMNIST dataset. In addition, on the CIFAR-10 dataset, where image classification is much harder than MNIST and FashionMNIST,  BMA-FL used only 14 aggregation rounds to reach the performance of FedAvg at 20th global aggregation round, which means BMA-FL is 30\% faster than FedAvg on CIFAR-10 dataset. Taking all three dataset into account, we can conclude that harder task will benefit more from BMA-FL achitecure. 

These comparative results underscore MASB-DRL's proficiency in expediting the federated learning process. By guiding aggregators to select the optimal local training frequency and local aggregation weights, MASB-DRL ensures that the trained local models exhibit reduced over-fitting and enhanced generalizability on previously unseen testing data.

\subsection{Blockchain Security Analysis}

\begin{table}[hbp!]
\centering
\caption{Times Edge Server Acts as Blockchain Miner with Malicious $P_1$}
\label{tab:robust}
\begin{tabular}{c|c}

\hline
Edge Servers  & \# of being a Miner \\ \hline
$P_1$ (Malicious) & 0.81                \\
$P_2$            & 8.73                \\
$P_3$             & 6.97                \\
$P_4$             & 9.32                \\
$P_5$            & 14.17               \\ \hline
\end{tabular}
\end{table}

We evaluate the robustness of proposed Performance-based Byzantine Consensus Mechanism (PBCM) in terms of the fault tolerance to malicious blockchain nodes. We make first edge server $P_1$ who is connected to two edge devices among the five edge servers as a malicious blockchain node, who does not train model or respond to blockchain consensus communication either. Table~\ref{tab:robust} shows the average count in 10 repeated execution that each edge server was selected as miner during 40 aggregations when BMA-FL conduct image classification on FashionMNIST dataset. We can see the malicious node $P_1$ mostly never been selected as miner, therefore the whole blockchain system will not be affected by this single malicious node. In fact, PBFT-based consensus mechanisms are proved to be secure if there are less than $\lfloor \frac{n-1}{3} \rfloor$ malicious blockchain nodes, where $n$ is the total number of blockchain nodes~\cite{DBLP:journals/tgcn/GuoDW22}.

\section{Conclusion}
\label{sec:conc}

This paper studied multi-aggregator federated learning problem in edge computing which is an under-explored topic. We proposed  Blockchain-empowered Heterogeneous Multi-Aggregator Federated Learning Architecture (BMA-FL) to fill the gap. BMA-FL addressed the challenge that The limited resources on edge devices make them not feasible for carrying the load for both model training and blockchain mining, which is often overlooked in related work. To this end, we proposed a Performance-based Byzantine Consensus Mechanism (PBCM) which is light-weight and secure for edge servers to choose honest one for global model aggregation. Due to the heterogeneity problem that edge devices have heterogeneous Non-IID data distribution and training speed, traditional model aggregation methods tend to underperform in the scenario. We proposed Multi-Agent with Shared Buffer Deep Reinforcement Learning algorithm (MASB-DRL) that enables edge servers learning the best training strategy from past experience. Empirical assessments indicated that BMA-FL surpasses benchmark models in terms of both training efficiency and end model performance, which demonstrates the effectiveness of PBCM and MASB-DRL. 

Our study does present areas for potential enhancement and further exploration. Firstly, data offloading from edge devices to servers could benefit from a refined strategy, determining consistent data quantities rather than the binary decision of uploading all or none of the untrained data each round. Secondly, while our approach to making the experience buffer shareable among edge servers involved integrating two server descriptive variables into the experience instance, this can be further optimized. Lastly, we still used traditional average strategy in global aggregation, which can also be further optimized with learnable aggregation strategies.




\bibliographystyle{IEEEtran}
\bibliography{ref}

\begin{thebibliography}{10}
\providecommand{\url}[1]{#1}
\csname url@samestyle\endcsname
\providecommand{\newblock}{\relax}
\providecommand{\bibinfo}[2]{#2}
\providecommand{\BIBentrySTDinterwordspacing}{\spaceskip=0pt\relax}
\providecommand{\BIBentryALTinterwordstretchfactor}{4}
\providecommand{\BIBentryALTinterwordspacing}{\spaceskip=\fontdimen2\font plus
\BIBentryALTinterwordstretchfactor\fontdimen3\font minus
  \fontdimen4\font\relax}
\providecommand{\BIBforeignlanguage}[2]{{%
\expandafter\ifx\csname l@#1\endcsname\relax
\typeout{** WARNING: IEEEtran.bst: No hyphenation pattern has been}%
\typeout{** loaded for the language `#1'. Using the pattern for}%
\typeout{** the default language instead.}%
\else
\language=\csname l@#1\endcsname
\fi
#2}}
\providecommand{\BIBdecl}{\relax}
\BIBdecl

\bibitem{DBLP:journals/iotj/SanghamiLH23}
\BIBentryALTinterwordspacing
S.~V. Sanghami, J.~J. Lee, and Q.~Hu, ``Machine-learning-enhanced blockchain
  consensus with transaction prioritization for smart cities,'' \emph{{IEEE}
  Internet Things J.}, vol.~10, no. 8, April 15, pp. 6661--6672, 2023.
  [Online]. Available: \url{https://doi.org/10.1109/JIOT.2022.3175208}
\BIBentrySTDinterwordspacing

\bibitem{DBLP:journals/iotj/NguyenDPPLSLNP21}
\BIBentryALTinterwordspacing
D.~C. Nguyen, M.~Ding, Q.~Pham, P.~N. Pathirana, L.~B. Le, A.~Seneviratne,
  J.~Li, D.~Niyato, and H.~V. Poor, ``Federated learning meets blockchain in
  edge computing: Opportunities and challenges,'' \emph{{IEEE} Internet Things
  J.}, vol.~8, no.~16, pp. 12\,806--12\,825, 2021. [Online]. Available:
  \url{https://doi.org/10.1109/JIOT.2021.3072611}
\BIBentrySTDinterwordspacing

\bibitem{DBLP:journals/comsur/NguyenCDLPLSLP21}
\BIBentryALTinterwordspacing
D.~C. Nguyen, P.~Cheng, M.~Ding, D.~L{\'{o}}pez{-}P{\'{e}}rez, P.~N. Pathirana,
  J.~Li, A.~Seneviratne, Y.~Li, and H.~V. Poor, ``Enabling {AI} in future
  wireless networks: {A} data life cycle perspective,'' \emph{{IEEE} Commun.
  Surv. Tutorials}, vol.~23, no.~1, pp. 553--595, 2021. [Online]. Available:
  \url{https://doi.org/10.1109/COMST.2020.3024783}
\BIBentrySTDinterwordspacing

\bibitem{DBLP:journals/iotj/LoLLWXPZ23}
\BIBentryALTinterwordspacing
S.~K. Lo, Y.~Liu, Q.~Lu, C.~Wang, X.~Xu, H.~Paik, and L.~Zhu, ``Toward
  trustworthy {AI:} blockchain-based architecture design for accountability and
  fairness of federated learning systems,'' \emph{{IEEE} Internet Things J.},
  vol.~10, no.~4, pp. 3276--3284, 2023. [Online]. Available:
  \url{https://doi.org/10.1109/JIOT.2022.3144450}
\BIBentrySTDinterwordspacing

\bibitem{DBLP:conf/aistats/McMahanMRHA17}
\BIBentryALTinterwordspacing
B.~McMahan, E.~Moore, D.~Ramage, S.~Hampson, and B.~A. y~Arcas,
  ``Communication-efficient learning of deep networks from decentralized
  data,'' in \emph{Proceedings of the 20th International Conference on
  Artificial Intelligence and Statistics, {AISTATS} 2017, 20-22 April 2017,
  Fort Lauderdale, FL, {USA}}, ser. Proceedings of Machine Learning Research,
  A.~Singh and X.~J. Zhu, Eds., vol.~54.\hskip 1em plus 0.5em minus 0.4em\relax
  {PMLR}, 2017, pp. 1273--1282. [Online]. Available:
  \url{http://proceedings.mlr.press/v54/mcmahan17a.html}
\BIBentrySTDinterwordspacing

\bibitem{DBLP:journals/kbs/ZhaoZC23}
\BIBentryALTinterwordspacing
Y.~Zhao, J.~Zhang, and Y.~Cao, ``Manipulating vulnerability: Poisoning attacks
  and countermeasures in federated cloud-edge-client learning for image
  classification,'' \emph{Knowl. Based Syst.}, vol. 259, p. 110072, 2023.
  [Online]. Available: \url{https://doi.org/10.1016/j.knosys.2022.110072}
\BIBentrySTDinterwordspacing

\bibitem{DBLP:conf/aaai/YanWYL23}
\BIBentryALTinterwordspacing
G.~Yan, H.~Wang, X.~Yuan, and J.~Li, ``Defl: Defending against model poisoning
  attacks in federated learning via critical learning periods awareness,'' in
  \emph{Thirty-Seventh {AAAI} Conference on Artificial Intelligence, {AAAI}
  2023, Thirty-Fifth Conference on Innovative Applications of Artificial
  Intelligence, {IAAI} 2023, Thirteenth Symposium on Educational Advances in
  Artificial Intelligence, {EAAI} 2023, Washington, DC, USA, February 7-14,
  2023}, B.~Williams, Y.~Chen, and J.~Neville, Eds.\hskip 1em plus 0.5em minus
  0.4em\relax {AAAI} Press, 2023, pp. 10\,711--10\,719. [Online]. Available:
  \url{https://doi.org/10.1609/aaai.v37i9.26271}
\BIBentrySTDinterwordspacing

\bibitem{DBLP:journals/iotj/ZhangGQXQW23}
\BIBentryALTinterwordspacing
F.~Zhang, S.~Guo, X.~Qiu, S.~Xu, F.~Qi, and Z.~Wang, ``Federated learning meets
  blockchain: State channel-based distributed data-sharing trust supervision
  mechanism,'' \emph{{IEEE} Internet Things J.}, vol.~10, no.~14, pp.
  12\,066--12\,076, 2023. [Online]. Available:
  \url{https://doi.org/10.1109/JIOT.2021.3130116}
\BIBentrySTDinterwordspacing

\bibitem{DBLP:journals/tpds/ShayanFYB21}
\BIBentryALTinterwordspacing
M.~Shayan, C.~Fung, C.~J.~M. Yoon, and I.~Beschastnikh, ``Biscotti: {A}
  blockchain system for private and secure federated learning,'' \emph{{IEEE}
  Trans. Parallel Distributed Syst.}, vol.~32, no.~7, pp. 1513--1525, 2021.
  [Online]. Available: \url{https://doi.org/10.1109/TPDS.2020.3044223}
\BIBentrySTDinterwordspacing

\bibitem{DBLP:journals/tii/LuHDMZ20a}
\BIBentryALTinterwordspacing
Y.~Lu, X.~Huang, Y.~Dai, S.~Maharjan, and Y.~Zhang, ``Blockchain and federated
  learning for privacy-preserved data sharing in industrial iot,'' \emph{{IEEE}
  Trans. Ind. Informatics}, vol.~16, no.~6, pp. 4177--4186, 2020. [Online].
  Available: \url{https://doi.org/10.1109/TII.2019.2942190}
\BIBentrySTDinterwordspacing

\bibitem{DBLP:journals/tpds/LiSWDMSHP22}
\BIBentryALTinterwordspacing
J.~Li, Y.~Shao, K.~Wei, M.~Ding, C.~Ma, L.~Shi, Z.~Han, and H.~V. Poor,
  ``Blockchain assisted decentralized federated learning {(BLADE-FL):}
  performance analysis and resource allocation,'' \emph{{IEEE} Trans. Parallel
  Distributed Syst.}, vol.~33, no.~10, pp. 2401--2415, 2022. [Online].
  Available: \url{https://doi.org/10.1109/TPDS.2021.3138848}
\BIBentrySTDinterwordspacing

\bibitem{DBLP:journals/tii/KalapaakingKRAY23}
\BIBentryALTinterwordspacing
A.~P. Kalapaaking, I.~Khalil, M.~S. Rahman, M.~Atiquzzaman, X.~Yi, and
  M.~Almashor, ``Blockchain-based federated learning with secure aggregation in
  trusted execution environment for internet-of-things,'' \emph{{IEEE} Trans.
  Ind. Informatics}, vol.~19, no.~2, pp. 1703--1714, 2023. [Online]. Available:
  \url{https://doi.org/10.1109/TII.2022.3170348}
\BIBentrySTDinterwordspacing

\bibitem{DBLP:journals/tpds/WangHLXX23}
\BIBentryALTinterwordspacing
Z.~Wang, Q.~Hu, R.~Li, M.~Xu, and Z.~Xiong, ``Incentive mechanism design for
  joint resource allocation in blockchain-based federated learning,''
  \emph{{IEEE} Trans. Parallel Distributed Syst.}, vol.~34, no.~5, pp.
  1536--1547, 2023. [Online]. Available:
  \url{https://doi.org/10.1109/TPDS.2023.3253604}
\BIBentrySTDinterwordspacing

\bibitem{DBLP:journals/ieeenl/HieuTNNKE22}
\BIBentryALTinterwordspacing
N.~Q. Hieu, T.~A. Tran, C.~L. Nguyen, D.~Niyato, D.~I. Kim, and E.~Elmroth,
  ``Deep reinforcement learning for resource management in blockchain-enabled
  federated learning network,'' \emph{{IEEE} Netw. Lett.}, vol.~4, no.~3, pp.
  137--141, 2022. [Online]. Available:
  \url{https://doi.org/10.1109/LNET.2022.3173971}
\BIBentrySTDinterwordspacing

\bibitem{DBLP:journals/tits/AloqailyRG22}
\BIBentryALTinterwordspacing
M.~Aloqaily, I.~A. Ridhawi, and M.~Guizani, ``Energy-aware blockchain and
  federated learning-supported vehicular networks,'' \emph{{IEEE} Trans.
  Intell. Transp. Syst.}, vol.~23, no.~11, pp. 22\,641--22\,652, 2022.
  [Online]. Available: \url{https://doi.org/10.1109/TITS.2021.3103645}
\BIBentrySTDinterwordspacing

\bibitem{DBLP:conf/bigdataconf/ChenJLLL18}
\BIBentryALTinterwordspacing
X.~Chen, J.~Ji, C.~Luo, W.~Liao, and P.~Li, ``When machine learning meets
  blockchain: {A} decentralized, privacy-preserving and secure design,'' in
  \emph{{IEEE} International Conference on Big Data {(IEEE} BigData 2018),
  Seattle, WA, USA, December 10-13, 2018}, N.~Abe, H.~Liu, C.~Pu, X.~Hu, N.~K.
  Ahmed, M.~Qiao, Y.~Song, D.~Kossmann, B.~Liu, K.~Lee, J.~Tang, J.~He, and
  J.~S. Saltz, Eds.\hskip 1em plus 0.5em minus 0.4em\relax {IEEE}, 2018, pp.
  1178--1187. [Online]. Available:
  \url{https://doi.org/10.1109/BigData.2018.8622598}
\BIBentrySTDinterwordspacing

\bibitem{DBLP:journals/iotj/QuGLXYLZ20}
\BIBentryALTinterwordspacing
Y.~Qu, L.~Gao, T.~H. Luan, Y.~Xiang, S.~Yu, B.~Li, and G.~Zheng,
  ``Decentralized privacy using blockchain-enabled federated learning in fog
  computing,'' \emph{{IEEE} Internet Things J.}, vol.~7, no.~6, pp. 5171--5183,
  2020. [Online]. Available: \url{https://doi.org/10.1109/JIOT.2020.2977383}
\BIBentrySTDinterwordspacing

\bibitem{DBLP:journals/icl/KimPBK20}
\BIBentryALTinterwordspacing
H.~Kim, J.~Park, M.~Bennis, and S.~Kim, ``Blockchained on-device federated
  learning,'' \emph{{IEEE} Commun. Lett.}, vol.~24, no.~6, pp. 1279--1283,
  2020. [Online]. Available: \url{https://doi.org/10.1109/LCOMM.2019.2921755}
\BIBentrySTDinterwordspacing

\bibitem{DBLP:journals/csur/IssaMTST23}
\BIBentryALTinterwordspacing
W.~Issa, N.~Moustafa, B.~P. Turnbull, N.~Sohrabi, and Z.~Tari,
  ``Blockchain-based federated learning for securing internet of things: {A}
  comprehensive survey,'' \emph{{ACM} Comput. Surv.}, vol.~55, no.~9, pp.
  191:1--191:43, 2023. [Online]. Available:
  \url{https://doi.org/10.1145/3560816}
\BIBentrySTDinterwordspacing

\bibitem{DBLP:journals/cn/WanQGX22}
\BIBentryALTinterwordspacing
Y.~Wan, Y.~Qu, L.~Gao, and Y.~Xiang, ``Privacy-preserving blockchain-enabled
  federated learning for b5g-driven edge computing,'' \emph{Comput. Networks},
  vol. 204, p. 108671, 2022. [Online]. Available:
  \url{https://doi.org/10.1016/j.comnet.2021.108671}
\BIBentrySTDinterwordspacing

\bibitem{DBLP:journals/iotj/HuWXC23}
\BIBentryALTinterwordspacing
Q.~Hu, Z.~Wang, M.~Xu, and X.~Cheng, ``Blockchain and federated edge learning
  for privacy-preserving mobile crowdsensing,'' \emph{{IEEE} Internet Things
  J.}, vol.~10, no.~14, pp. 12\,000--12\,011, 2023. [Online]. Available:
  \url{https://doi.org/10.1109/JIOT.2021.3128155}
\BIBentrySTDinterwordspacing

\bibitem{DBLP:journals/cn/WangWHMSTC22}
\BIBentryALTinterwordspacing
W.~Wang, Y.~Wang, Y.~Huang, C.~Mu, Z.~Sun, X.~Tong, and Z.~Cai, ``Privacy
  protection federated learning system based on blockchain and edge computing
  in mobile crowdsourcing,'' \emph{Comput. Networks}, vol. 215, p. 109206,
  2022. [Online]. Available: \url{https://doi.org/10.1016/j.comnet.2022.109206}
\BIBentrySTDinterwordspacing

\bibitem{DBLP:journals/iotj/OtoumRM23}
\BIBentryALTinterwordspacing
S.~Otoum, I.~A. Ridhawi, and H.~T. Mouftah, ``A federated learning and
  blockchain-enabled sustainable energy trade at the edge: {A} framework for
  industry 4.0,'' \emph{{IEEE} Internet Things J.}, vol.~10, no.~4, pp.
  3018--3026, 2023. [Online]. Available:
  \url{https://doi.org/10.1109/JIOT.2022.3140430}
\BIBentrySTDinterwordspacing

\bibitem{DBLP:journals/iotj/JiangZTXZ22}
\BIBentryALTinterwordspacing
L.~Jiang, H.~Zheng, H.~Tian, S.~Xie, and Y.~Zhang, ``Cooperative federated
  learning and model update verification in blockchain-empowered digital twin
  edge networks,'' \emph{{IEEE} Internet Things J.}, vol.~9, no.~13, pp.
  11\,154--11\,167, 2022. [Online]. Available:
  \url{https://doi.org/10.1109/JIOT.2021.3126207}
\BIBentrySTDinterwordspacing

\bibitem{DBLP:journals/iotj/LuHZMZ21}
\BIBentryALTinterwordspacing
Y.~Lu, X.~Huang, K.~Zhang, S.~Maharjan, and Y.~Zhang, ``Communication-efficient
  federated learning and permissioned blockchain for digital twin edge
  networks,'' \emph{{IEEE} Internet Things J.}, vol.~8, no.~4, pp. 2276--2288,
  2021. [Online]. Available: \url{https://doi.org/10.1109/JIOT.2020.3015772}
\BIBentrySTDinterwordspacing

\bibitem{DBLP:journals/iotj/CuiSMCYZX22}
\BIBentryALTinterwordspacing
L.~Cui, X.~Su, Z.~Ming, Z.~Chen, S.~Yang, Y.~Zhou, and W.~Xiao, ``{CREAT:}
  blockchain-assisted compression algorithm of federated learning for content
  caching in edge computing,'' \emph{{IEEE} Internet Things J.}, vol.~9,
  no.~16, pp. 14\,151--14\,161, 2022. [Online]. Available:
  \url{https://doi.org/10.1109/JIOT.2020.3014370}
\BIBentrySTDinterwordspacing

\bibitem{DBLP:journals/peerj-cs/MengL22}
\BIBentryALTinterwordspacing
M.~Meng and Y.~Li, ``Sfedchain: blockchain-based federated learning scheme for
  secure data sharing in distributed energy storage networks,'' \emph{PeerJ
  Comput. Sci.}, vol.~8, p. e1027, 2022. [Online]. Available:
  \url{https://doi.org/10.7717/peerj-cs.1027}
\BIBentrySTDinterwordspacing

\bibitem{DBLP:journals/iotj/FanZZC21}
\BIBentryALTinterwordspacing
S.~Fan, H.~Zhang, Y.~Zeng, and W.~Cai, ``Hybrid blockchain-based resource
  trading system for federated learning in edge computing,'' \emph{{IEEE}
  Internet Things J.}, vol.~8, no.~4, pp. 2252--2264, 2021. [Online].
  Available: \url{https://doi.org/10.1109/JIOT.2020.3028101}
\BIBentrySTDinterwordspacing

\bibitem{DBLP:journals/tvt/LiuZZZSPZ21}
\BIBentryALTinterwordspacing
H.~Liu, S.~Zhang, P.~Zhang, X.~Zhou, X.~Shao, G.~Pu, and Y.~Zhang, ``Blockchain
  and federated learning for collaborative intrusion detection in vehicular
  edge computing,'' \emph{{IEEE} Trans. Veh. Technol.}, vol.~70, no.~6, pp.
  6073--6084, 2021. [Online]. Available:
  \url{https://doi.org/10.1109/TVT.2021.3076780}
\BIBentrySTDinterwordspacing

\bibitem{DBLP:journals/jsac/NguyenHLPB22}
\BIBentryALTinterwordspacing
D.~C. Nguyen, S.~Hosseinalipour, D.~J. Love, P.~N. Pathirana, and C.~G.
  Brinton, ``Latency optimization for blockchain-empowered federated learning
  in multi-server edge computing,'' \emph{{IEEE} J. Sel. Areas Commun.},
  vol.~40, no.~12, pp. 3373--3390, 2022. [Online]. Available:
  \url{https://doi.org/10.1109/JSAC.2022.3213344}
\BIBentrySTDinterwordspacing

\bibitem{DBLP:journals/sensors/LeeK22b}
\BIBentryALTinterwordspacing
J.~Lee and W.~Kim, ``Dag-based blockchain sharding for secure federated
  learning with non-iid data,'' \emph{Sensors}, vol.~22, no.~21, p. 8263, 2022.
  [Online]. Available: \url{https://doi.org/10.3390/s22218263}
\BIBentrySTDinterwordspacing

\bibitem{DBLP:journals/tpds/QuWHC21}
\BIBentryALTinterwordspacing
X.~Qu, S.~Wang, Q.~Hu, and X.~Cheng, ``Proof of federated learning: {A} novel
  energy-recycling consensus algorithm,'' \emph{{IEEE} Trans. Parallel
  Distributed Syst.}, vol.~32, no.~8, pp. 2074--2085, 2021. [Online].
  Available: \url{https://doi.org/10.1109/TPDS.2021.3056773}
\BIBentrySTDinterwordspacing

\bibitem{DBLP:journals/jsac/WangPSLBW22}
\BIBentryALTinterwordspacing
Y.~Wang, H.~Peng, Z.~Su, T.~H. Luan, A.~Benslimane, and Y.~Wu, ``A
  platform-free proof of federated learning consensus mechanism for sustainable
  blockchains,'' \emph{{IEEE} J. Sel. Areas Commun.}, vol.~40, no.~12, pp.
  3305--3324, 2022. [Online]. Available:
  \url{https://doi.org/10.1109/JSAC.2022.3213347}
\BIBentrySTDinterwordspacing

\bibitem{DBLP:journals/network/LiCLHZY21}
\BIBentryALTinterwordspacing
Y.~Li, C.~Chen, N.~Liu, H.~Huang, Z.~Zheng, and Q.~Yan, ``A blockchain-based
  decentralized federated learning framework with committee consensus,''
  \emph{{IEEE} Netw.}, vol.~35, no.~1, pp. 234--241, 2021. [Online]. Available:
  \url{https://doi.org/10.1109/MNET.011.2000263}
\BIBentrySTDinterwordspacing

\bibitem{DBLP:journals/iotj/ChenLWYLXZC23}
\BIBentryALTinterwordspacing
Y.~Chen, J.~Li, F.~Wang, K.~Yue, Y.~Li, B.~Xing, L.~Zhang, and L.~Chen,
  ``{DS2PM:} {A} data-sharing privacy protection model based on blockchain and
  federated learning,'' \emph{{IEEE} Internet Things J.}, vol.~10, no.~14, pp.
  12\,112--12\,125, 2023. [Online]. Available:
  \url{https://doi.org/10.1109/JIOT.2021.3134755}
\BIBentrySTDinterwordspacing

\bibitem{10159403}
Y.~Lin, Z.~Gao, H.~Du, J.~Kang, D.~Niyato, Q.~Wang, J.~Ruan, and S.~Wan,
  ``Drl-based adaptive sharding for blockchain-based federated learning,''
  \emph{IEEE Transactions on Communications}, pp. 1--1, 2023.

\bibitem{DBLP:journals/tvt/LuHZMZ20}
\BIBentryALTinterwordspacing
Y.~Lu, X.~Huang, K.~Zhang, S.~Maharjan, and Y.~Zhang, ``Blockchain empowered
  asynchronous federated learning for secure data sharing in internet of
  vehicles,'' \emph{{IEEE} Trans. Veh. Technol.}, vol.~69, no.~4, pp.
  4298--4311, 2020. [Online]. Available:
  \url{https://doi.org/10.1109/TVT.2020.2973651}
\BIBentrySTDinterwordspacing

\bibitem{castro1999practical}
M.~Castro, B.~Liskov \emph{et~al.}, ``Practical byzantine fault tolerance,'' in
  \emph{OsDI}, vol.~99, no. 1999, 1999, pp. 173--186.

\bibitem{DBLP:journals/tgcn/GuoDW22}
\BIBentryALTinterwordspacing
J.~Guo, X.~Ding, and W.~Wu, ``An architecture for distributed energies trading
  in byzantine-based blockchains,'' \emph{{IEEE} Trans. Green Commun. Netw.},
  vol.~6, no.~2, pp. 1216--1230, 2022. [Online]. Available:
  \url{https://doi.org/10.1109/TGCN.2022.3142438}
\BIBentrySTDinterwordspacing

\bibitem{DBLP:journals/corr/LillicrapHPHETS15}
\BIBentryALTinterwordspacing
T.~P. Lillicrap, J.~J. Hunt, A.~Pritzel, N.~Heess, T.~Erez, Y.~Tassa,
  D.~Silver, and D.~Wierstra, ``Continuous control with deep reinforcement
  learning,'' in \emph{4th International Conference on Learning
  Representations, {ICLR} 2016, San Juan, Puerto Rico, May 2-4, 2016,
  Conference Track Proceedings}, Y.~Bengio and Y.~LeCun, Eds., 2016. [Online].
  Available: \url{http://arxiv.org/abs/1509.02971}
\BIBentrySTDinterwordspacing

\bibitem{lecun2010mnist}
Y.~LeCun, C.~Cortes, and C.~Burges, ``Mnist handwritten digit database,''
  \emph{ATT Labs [Online]. Available: http://yann.lecun.com/exdb/mnist},
  vol.~2, 2010.

\bibitem{DBLP:journals/corr/abs-1708-07747}
\BIBentryALTinterwordspacing
H.~Xiao, K.~Rasul, and R.~Vollgraf, ``Fashion-mnist: a novel image dataset for
  benchmarking machine learning algorithms,'' \emph{CoRR}, vol. abs/1708.07747,
  2017. [Online]. Available: \url{http://arxiv.org/abs/1708.07747}
\BIBentrySTDinterwordspacing

\bibitem{Krizhevsky09learningmultiple}
A.~Krizhevsky, ``Learning multiple layers of features from tiny images,'' Tech.
  Rep., 2009.

\end{thebibliography}


\begin{IEEEbiography}[{\includegraphics[width=1in,height=1.25in,clip,keepaspectratio]{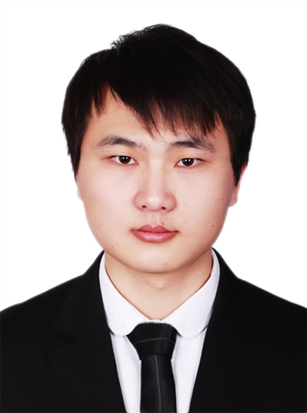}}]{Xiao Li} (Student Member, IEEE) 
received his B.S. and M.S degree in Software Engineering from
Dalian University of Technology, China in 2016 and 2019, respectively. He is
currently pursuing the Ph.D. degree with the Department of Computer Science,
The University of Texas at Dallas, Richardson, TX, USA. His current research
interests include data mining, machine learning, distributed systems and blockchain.
\end{IEEEbiography}

\begin{IEEEbiography}[{\includegraphics[width=1in,height=1.25in,clip,keepaspectratio]{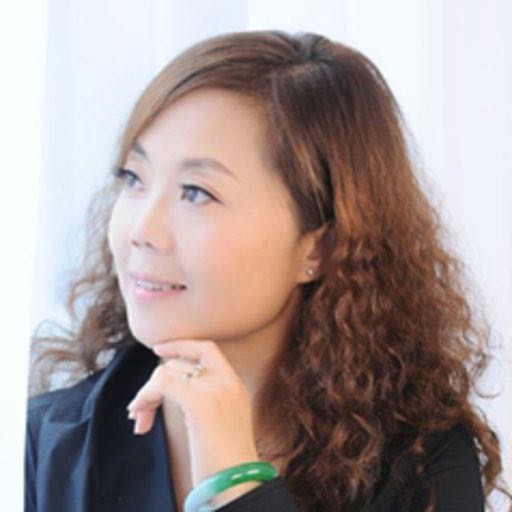}}]{Weili Wu}
(Senior Member, IEEE) received the M.S.
and Ph.D. degrees from the Department of Computer
Science, University of Minnesota, Minneapolis, MN,
USA, in 1998 and 2002, respectively.
She is currently a Full Professor with the
Department of Computer Science, The University of
Texas at Dallas, Richardson, TX, USA. Her research
mainly deals in the general research area of data
communication and data management. Her research
focuses on the design and analysis of algorithms
for optimization problems that occur in wireless
networking environments and various database systems.
\end{IEEEbiography}


\end{document}